\documentclass[12pt]{article}
\usepackage{amsmath}
\usepackage{graphicx}%
\usepackage{amsfonts}%
\usepackage{amssymb}
\usepackage{enumerate}
\usepackage{natbib}
\usepackage{url} 
\usepackage{color}
\usepackage{multicol}
\usepackage{multirow}
\usepackage{booktabs}
\usepackage{xurl}

\usepackage{float}
\restylefloat{figure}

\usepackage[left]{lineno}

\newcommand{\blind}{1}

\newcommand{\bbeta}{ \mbox{\boldmath $ \beta $} }

\newcommand{\btheta}{ \mbox{\boldmath $ \theta $} }

\newcommand{\bdelta}{ \mbox{\boldmath $\delta$} }

\newcommand{\bx}{\textbf{x}}

\begin{document}

\def\spacingset#1{\renewcommand{\baselinestretch}%
{#1}\small\normalsize} \spacingset{1}


\if1\blind
{
  \title{
  Analyzing whale calling through Hawkes process modeling}
  
  \author{Bokgyeong Kang\\
    Department of Statistical Science, Duke University,\\
    \\
    Erin M. Schliep\\
    Department of Statistics, North Carolina State University,\\
    \\
    Alan E. Gelfand\\
    Department of Statistical Science, Duke University,\\
    \\
    Tina M. Yack\\
    Nicholas School of the Environment, Duke University,\\
    \\
    Christopher W. Clark\\
    K. Lisa Yang Center for Conservation Bioacoustics, Cornell University,\\
    and \\
    Robert S. Schick\\
    Southall Environmental Associates
    }
  \maketitle
} \fi

\if0\blind
{
  \bigskip
  \bigskip
  \bigskip
  \begin{center}
    {\LARGE\bf Analyzing whale calling through Hawkes process modeling}
\end{center}
  \medskip
} \fi

\begin{abstract}
Sound is assumed to be the primary modality of communication among marine mammal species. Analyzing acoustic recordings helps to understand the function of the acoustic signals as well as the possible impact of anthropogenic noise on acoustic behavior. Motivated by a dataset from a network of hydrophones in Cape Cod Bay, Massachusetts, utilizing automatically detected calls in recordings, we study the communication process of the endangered North Atlantic right whale. For right whales an ``up-call'' is known as a contact call, and ensuing counter-calling between individuals is presumed to facilitate group cohesion.  We present novel spatiotemporal excitement modeling consisting of a \textit{background} process and a \textit{counter-call} process. The background process intensity incorporates the influences of diel patterns and ambient noise on occurrence. The counter-call intensity captures potential excitement, that calling elicits calling behavior. Call incidence is found to be clustered in space and time; a call seems to excite more calls nearer to it in time and space. We find evidence that whales make more calls during twilight hours, respond to other whales nearby, and are likely to remain quiet in the presence of increased ambient noise.

\end{abstract}


\noindent%

\spacingset{1.8} 




\section{Introduction}
\label{sec:intro}

Sound is assumed to be the primary mode of communication among many marine mammal species \citep{bass2003physical}. Scientists have been recording the sounds produced by marine mammals for many decades \citep{schevill1962photograph,payne1971songs,tyack2000communication}, and over time have developed more advanced methods to detect \citep{mooreListeningLargeWhales2006,mellingerOverviewFixedPassive2007}, classify \citep{gillespie2004detection}, localize \citep{watkinsSoundSourceLocation1972a}, and understand the utility of acoustic signals produced by these animals \citep{richardson1995marine, au2008principles, bradbury1998principles}. 

Our focus is on the North Atlantic right whale (\textit{Eubalaena glacialis}, hereafter NARW), an endangered baleen whale whose primary habitat is the western North Atlantic Ocean \citep{krausUrbanWhaleNorth2007}. NARWs face a variety of threats, including ship strikes \citep{vanderlaan2007}, entanglement with fishing gear \citep{knowltonFishingGearEntanglement2022}, and climate-change induced impacts on their habitat  \citep{meyer-gutbrodRedefiningNorthAtlantic2023}. Their population is currently estimated to be 356 (346-363) animals \citep{linden2023}. NARWs have a diverse acoustic repertoire \citep{clark1982,matthewsOverviewNorthAtlantic2021}, with a variety
of different call types that vary by habitat \citep{cusanoImplementingConservationMeasures2019}, behavior \citep{parks2005sound} and age \citep{root-gutteridgeLifetimeChangingCalls2018}. One of these call types is a frequency-modulated ``up-call'' \citep{clark1982}, which is thought to serve as a \emph{contact} call between individual whales \citep{clark1980sound}.  The ensuing \emph{counter-calling} between individuals, in response to the up-calls, is presumed to serve as a mechanism for maintaining group cohesion. Up-calls have been used to document the presence and relative abundance of NARWs across broad spatial \citep{davisLongtermPassiveAcoustic2017} and temporal scales \citep{davisUpcallingBehaviourPatterns2023}. Our analysis aims to better understand the dynamics of up-calling behavior of NARWs within and across days. 


Tracking the calls of individuals through time is challenging. It is often done with a combination of survey platforms \citep{clark1986preliminary}, on navy ranges with a dense array of hydrophones \citep{helble2016swim}, or using short-term movement and acoustic tags on animals \citep{johnsonDigitalAcousticRecording2003,parksIndividualRightWhales2010,parksAcousticCrypsisCommunication2019}. Though movement and acoustic tags allow you to detect the precise timing of calling by the tagged individual, they are not without limitations; for example, 1) the tag deployments are short-term, typically on the order of hours; 2) they are archival, so they must be retrieved; and 3), as compared to an array of hydrophones, they are spatially limited. In contrast, a distributed array of acoustic recorders can cover a broader spatial area over a longer period of time. The acoustic recorder data analyzed here also have limitations, including the fact that they preclude direct observation of up-calls from known individuals. These data consist of the set of times at which individual NARW up-calls were received at each acoustic recorder in a distributed array of 10 recorders. Each recorder captures a unique sequence of up-call detection times. At the level of the array, each sequence is a marked point pattern, i.e., an event time with an associated recorder label for each fixed recorder location.


Our analysis aims to better understand how diel patterns and ambient noise levels influence up-calling behavior across individuals \citep{matthewsOverviewNorthAtlantic2021}. We also seek to identify evidence of counter-calling behavior, i.e., whether NARWs produce up-calls in response to other up-calling whales. If counter-calling behavior is present, the aggregated sequence will exhibit excitement since the incidence of an up-call tends to encourage \textit{more} up-calls than otherwise expected in a window of time after the up-call. This excitement could occur both within or across recorders. Lastly, given concerns about levels of anthropogenic noise in the ocean \citep{radfordAcousticCommunicationNoisy2014,hatchCanYouHear2016}, there is a need to quantify the potential impacts of anthropogenic sounds on the calling behavior of NARWs. To accomplish these goals, we specify a novel multivariate Hawkes process where the intensity for the up-calls is the summation of a so-called background process that produces contact calls and a counter-call process that can explain the potential excitement.\footnote{The term intensity has very different meaning for acousticians, who use it to describe the strength or amplitude of a sound. Throughout the manuscript, we use intensity in the statistical sense.} The notion of cross-recorder excitement can potentially provide insight into communication between different individuals. Very recent ecological work \citep{Nicvert2024} has employed a multivariate Hawkes process to study interactions between multiple species from camera trap data across time.

Since each acoustic sensor has a unique spatial location, joint modeling of the event sequences from the recorders requires spatial specification. We assume that neighboring recorders exhibit similar patterns of contact calling. Similarly, we assume that an up-call first detected at a particular recorder is more likely to excite counter-calls at the same or nearby recorders. As such, a model for multi-recorder data requires the introduction of spatial dependence. We introduce spatial dependence into the background process through spatially-varying regression coefficients. We introduce distance-based dependence into the counter-call (CC) process. Further, we expand the customary specification for the multivariate Hawkes process by enriching the background process from a nonhomogeneous Poisson process (NHPP) to a process allowing for local adjustment to the intensity through a Gaussian process (GP). Altogether, we label this model as NHPP+GP+CC. This is our richest specification and we investigate it with regard to its ability to distinguish between the GP and CC components of the overall intensity.

The models developed here differ from customary spatial self-exciting process modeling in the literature \citep[see, e.g.,][]{Rathbun1996,Reinhart2018}. In the usual setting, the data consist of a marked point pattern---a single event time sequence with a mark denoting the spatial location over the region. This formulation results in a conditional intensity over space and time. In our analysis, we specify a conditional intensity of the event time sequence for each hydrophone, reflecting its spatial location. These conditional intensities are modeled jointly across space. Conceptually similar is the network Hawkes modeling in \citet{lindermanDiscoveringLatentNetwork2014}.

We consider the following ensemble of models: (i) NHPP; (ii) NHPP+GP; (iii) NHPP+CC; (iv) NHPP+GP+CC. Evidently, models (i) to (iii) are special cases of (iv). All model specification and fitting are done in a hierarchical Bayesian framework. We introduce a simulation study in order to learn how well the fitting models can distinguish generating models. To assess model adequacy, we extend the random time change theorem (RTCT)
\citep{Brown1988} to a fully Bayesian modeling setting. For model comparison, we add posterior likelihood comparison along with model dimension penalization through the Deviance Information Criterion (DIC).

Our primary contribution is an analysis of two NARW vocalization datasets, one from a single recorder located just west of Race Point at the Northern edge of Cape Cod Bay and the second from a network of recorders within the central part of Cape Cod Bay (CCB). We employ the event series of received up-call times to explore three features: 1) the periodic/diel behavior of contact calls; 2) evidence of counter-calling behavior in the data; and 3) the role of ambient noise in altering the up-calling behavior in NARW.

The format of the paper is as follows. Section \ref{sec:data} provides details of the data collection as well as some exploratory data analysis for the single and multi-channel data sources. Section \ref{sec:models} elaborates the multi-channel models with the single channel model as a special case.  Section \ref{sec:assess} introduces model adequacy and model comparison tools. Section \ref{sec:sim} offers some simulation work to demonstrate the ability of our modeling to recover the true generative specification.  Section \ref{sec:real} provides our findings for both the single channel and the multi-channel.  Section \ref{sec:summary} concludes with a summary of our contributions, our findings, and some plans for future work.

\section{Passive acoustic monitoring data}
\label{sec:data}

Researchers have been monitoring and recording the sounds produced by NARWs for several decades, and acoustic detections have provided information about the spatial and temporal distribution of these whales  \citep{davisLongtermPassiveAcoustic2017}. 
We focus on one type of sound referred to as the up-call, which serves as a contact call to maintain social cohesion \citep{clark1982}. Acoustic data have been collected with a distributed array of bottom-mounted hydrophones which remain underwater for many months at a time while continuously recording acoustic data \citep{calupca2000}. From these continuous records, bioacousticians use a variety of methods to detect and classify the call made by individual(s) \citep{clark1980sound}. This process leads to a temporally explicit series of unique up-call events. We explore up-calling behavior of NARWs recorded from both a single-channel recorder and a distributed network of 10 recorders.

\subsection{Single-channel data}
\label{subsec:sdata}

\begin{figure}[!t]
\begin{center}
\includegraphics[width = 0.73\textwidth]{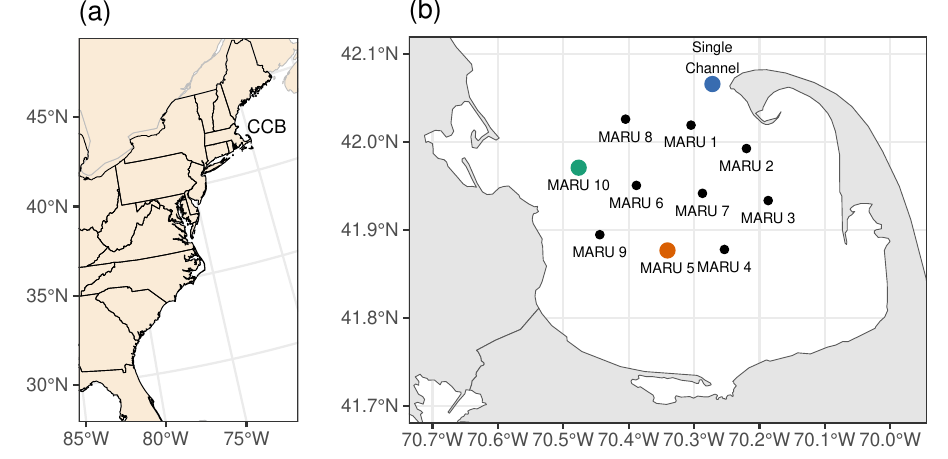}
\end{center}
\caption{(a) The location of Cape Cod Bay, MA. (b) The locations of the single MARU (blue circle) and the distributed network of 10 MARUs in CCB. The orange and green colors correspond to two highlighted MARUs which will be referred to in subsequent figures.
\label{fig:ccbHP}}
\end{figure}

The single-channel dataset was collected and processed by the National Oceanic and Atmospheric Administration (NOAA) \citep{dclde2023}. 
This comprised recordings from a week-long deployment of a bottom-mounted hydrophone system, referred to as a marine autonomous recording unit or MARU \citep{calupca2000}; the MARU was deployed to the west of Race Point on the northern edge of Cape Cod Bay (Figure~\ref{fig:ccbHP}). MARUs were programmed to record sound continuously in the 10-585 Hz low-frequency band \citep{parksVariabilityAmbientNoise2009} including biotic, abiotic, and anthropogenic sources. Post-processing was done with a software routine designed to automatically detect and classify low-frequency calls of baleen whales, including NARW \citep{baumgartnerGeneralizedBaleenWhale2011}. Following automated detection, a trained analyst manually reviewed and verified possible NARW up-calls, and the processed dataset was made available to the acoustic processing community as a benchmark dataset to be used in subsequent algorithm development. There were 5,302 unique and verified up-calls detected across the seven days (2009-03-28 to 2009-04-03), and we extracted ambient noise levels (dB re 1 µPa) across the same frequency band (60-400 Hz) as the NARW up-call using the PAMGuard software program (https://www.pamguard.org). Ambient sound was in dB re 1 \textmu-Pa, i.e., decibel units of Sound Pressure Level, root mean square, with reference to 1 \textmu-Pascal---a standard metric in the bioacoustics community \citep{au2008principles}. This noise metric was used as a covariate in the modeling.

\subsection{Multi-channel CCB data}
\label{subsec:mdata}

\begin{figure}[!b]
\begin{center}
\includegraphics[width = 0.8\textwidth]{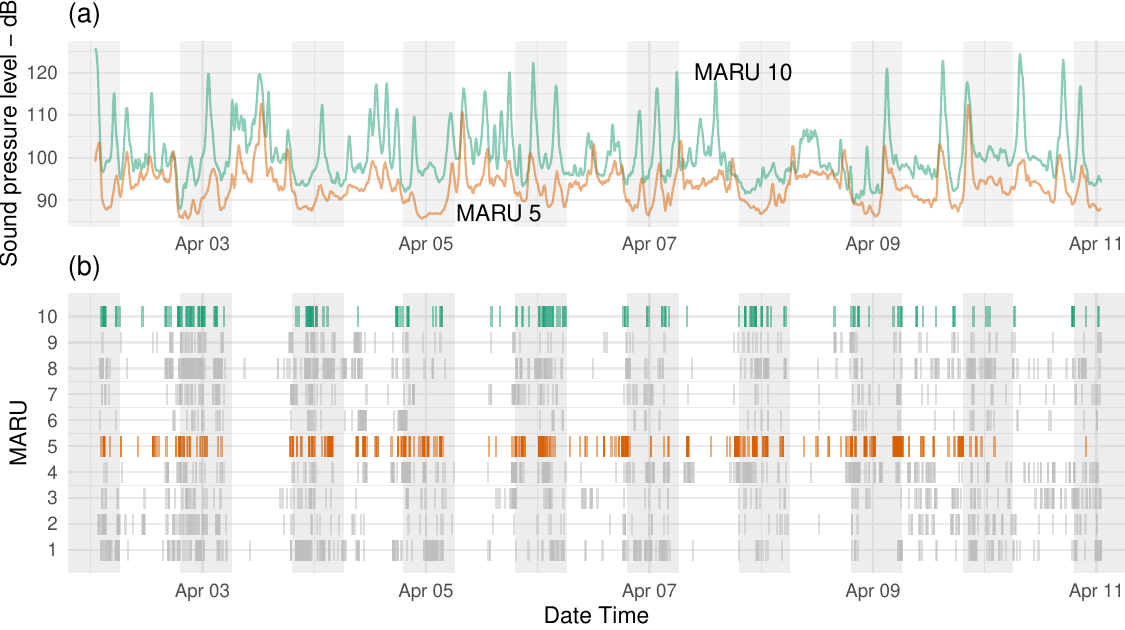}
\end{center}
\caption{(a) 30-minute rolling average time series of ambient noise measured at two example MARUs in CCB (MARU 10 in green; MARU 5 in orange). Units are decibels, Root-Mean-Squared. (b) Event sequence of unique up-calls received for each MARU within CCB, colors as in (a).  \label{fig:noiseCCB}}
\end{figure}

The multi-channel dataset consists of recordings from the distributed network of 10 MARUs located within Cape Cod Bay, MA. 
The distances between MARUs are found in the supplement Figure~S1. 
This bay is seasonally inhabited by NARWs \citep{mayoDistributionDemographyBehavior2018}, with upwards of 70\% of the entire species sometimes being observed during aerial surveys and vessel-based surveys (C. Hudak, pers comm). Starting in the early 2000s, researchers  deployed passive acoustic arrays of different size and configuration in CCB coincident with the season when NARWs occur in this habitat \citep{clarkVisualAcousticSurveys2010a,urazghildiievStatisticalAnalysisNorth2014}. These arrays use the same type of MARU used in the single-channel dataset described earlier, however the data processing is slightly different. For our multi-channel analysis, we used acoustic data collected from 10 MARUs in a distributed network across 9 days (2010-04-02 to 2010-04-10). The data are archival, so the buoys must be recovered to obtain the sound files \citep{calupca2000}. 

\begin{figure}[!b]
\begin{center}
\includegraphics[width = 0.65\textwidth]{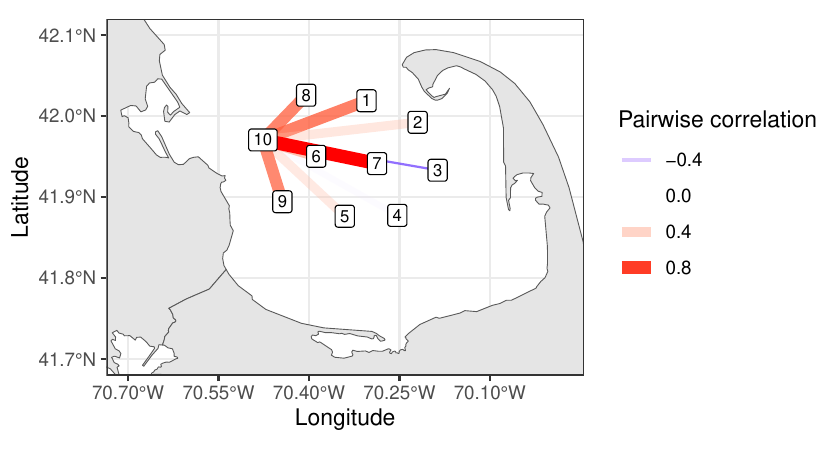}
\end{center}
\caption{Pairwise correlation between a kernel intensity estimate for the total event sequence at MARU 10 versus the kernel intensity estimates for each of the other nine MARUs. \label{fig:ccbCor10}}
\end{figure}

We used the software PAMGuard to process the raw sound files, detect and classify NARW up-calls, extract ambient noise metrics, and localize individual up-calling NARWs \citep{gillespie2004detection,gillespiePassiveAcousticMethods2020}. Localization algorithms use the time-difference-of-arrival of three or more matched up-calls \citep{urazghildiievStatisticalAnalysisNorth2014}. To determine the times of unique up-calls, i.e., those arriving from a single whale, we found all up-calls associated with each unique localization; from this associated set we extracted the first arrival time, as well as its associated MARU. Across the 9 days, a total of 2,750 unique up-calls were extracted from the data. The arrival times on the MARUs are the data that comprise the analysis. Figure~\ref{fig:noiseCCB} (a) shows the ambient noise for two illustrative MARUs (MARU 5 and MARU 10 in Figure \ref{fig:ccbHP})) highlighting variation across MARUs as well as within MARUs across days.
Figure~\ref{fig:noiseCCB} (b) depicts the event sequence of unique up-calls received on each of the MARUs within CCB for the 8-day period.
The up-call sequences show diel patterns as well as possible excitement in up-call rates, motivating the inclusion of diel harmonics in the intensity of the process as well as Hawkes process modeling. 

Finally, we investigate possible spatial dependence in the up-calling behavior captured by the array of MARUs. We might expect nearby hydrophones to capture similar up-calling behavior as a response to noise and temporal harmonics. In addition, we might expect the number of counter-calls received at MARU $k$ that were excited by up-calls at MARU $l$ to decrease with the distance between MARUs $l$ and $k$. For each MARU, we compute the kernel intensity estimate for the total event sequence of up-calls. Then, we compute pairwise correlations between these estimates. In Figure~\ref{fig:ccbCor10}, we show these pairwise correlations spatially between MARU 10 and each of the other MARUs. Positive correlation is detected between MARU 10 and each of the nearby MARUs, suggesting evidence of spatial dependence across the array.



\section{Model specification and inference}
\label{sec:models}

\subsection{Multivariate Hawkes process}
\label{subsec:models}
We propose four different multivariate Hawkes process models for studying the behavior of NARW up-calls.
Let $\mathcal{T} = \{ (t_i, m_i), i=1, \dots, n\}$ be the sequence of paired observations where $t_i \in (0, T]$ is an up-call time with $t_i < t_{i+1}$ and $m_i \in \{1, \dots, K\}$ indicates the MARU which received up-call $i$ at time $t_i$. Let $\mathcal{H}_{t}$ be the history of all up-calls up to time $t$, i.e., $\mathcal{H}_{t} = \{(t_i, m_i): t_i <t\}$. Such a point process in time can be characterized by a conditional intensity defined as
\begin{align*}
    \lambda(t\mid \mathcal{H}_{t}) = \lim_{\Delta t \to 0} \frac{ \textup{E}\left\{ N \left((t, t + \Delta t ] \right) \mid \mathcal{H}_{t} \right\} }{\Delta t },
\end{align*}
where $N(A)$ is the number of up-calls recorded over $A \subseteq (0, T]$. 

We label our most general model as NHPP+GP+CC (nonhomogeneous Poisson process + Gaussian process + counter-call process). For this model we define the conditional intensity of up-calls at MARU $k$ at time $t$ as
\begin{align}
    \lambda_k(t \mid \mathcal{H}_t; \btheta) &= \mu_k(t) + \sum_{i:t_i < t, t_i \in \mathcal{T}} \alpha_{m_i} e^{-\eta(t-t_i)} e^{-\phi d_{m_i,k}} \label{eq:multi_mukt},
\end{align}
where $\btheta$ is the collection of model parameters. The conditional intensity of up-calls across $K$ MARUs at time $t$ is $\lambda(t \mid \mathcal{H}_t; \btheta) = \sum_{k=1}^K \lambda_k(t \mid \mathcal{H}_t; \btheta)$. 

The first term $\mu_k(t)$ of \eqref{eq:multi_mukt} explains the rate of contact calls at MARU $k$ at time $t$.
We use minutes and kilometers (km) for the units of time and space.
To capture the impact of noise on contact calls as well as the potential daily pattern of contact calls, we assume
\begin{align}
    \mu_k(t) = \exp\bigg\{ &\beta_{0k} + 
    \beta_{1k} \text{Noise}_k(t) + 
    \beta_{2k} \sin\left(\frac{2 \pi t}{8 \times 60 }\right) + 
    \beta_{3k} \cos\left(\frac{2 \pi t}{8 \times 60}\right) + 
    \beta_{4k} \sin\left(\frac{2 \pi t}{12 \times 60}\right) \nonumber \\
    & + \beta_{5k} \cos\left(\frac{2 \pi t}{12 \times 60}\right) + 
    \beta_{6k} \sin\left(\frac{2 \pi t}{24 \times 60}\right) + 
    \beta_{7k} \cos\left(\frac{2 \pi t}{24 \times 60}\right) + \delta_k w(t)
    \bigg\}, \nonumber
\end{align}
where Noise$_k(t)$ is the ambient noise level at MARU $k$ at time $t$, and the three sine and cosine pairs are the 8-hour, 12-hour, and 24-hour harmonics, respectively. The GP, $w(\cdot)$, is mean 0 and variance 1, and is used to account for variability in the contact call process not captured by the other covariates. We assume an exponential correlation function with an effective range of 3 hours to capture local temporal dependence. The $\delta_k$'s are MARU-specific coefficients for the GP. 
Let $\bbeta_j = \left( \beta_{j1}, \dots, \beta_{jK} \right)^\top$ denote the length $K$ variable-specific vector of coefficients, where $j=0$ denotes the intercept. Let $\bdelta = \left( \delta_1, \dots, \delta_K \right)^\top$ denote a vector of the GP coefficients. 
To account for spatial dependence, we assume that $\beta_{jk}$'s are dependent across $k$. Let $d_{k,\ell}$ denote the distance between MARUs $k$ and $\ell$. To improve Markov chain mixing, we incorporate customary hierarchical centering \citep{Gelfand1996}. For $j = 0, 1, \dots, p$, we specify $\bbeta_j \sim \text{MVN}(\Tilde{\beta}_j \textbf{1}, \tau_j V)$ where $V_{k, \ell}$ is $e^{-3 d_{k,\ell}/ \max_{k^\prime,\ell^\prime}\{ d_{k^\prime,\ell^\prime} \} }$ so that its effective range entirely covers the array of MARUs. The GP coefficients $\delta_k$'s are also assumed to be spatially dependent and given by $\log(\bdelta) \sim \text{MVN}(\Tilde{\delta} \textbf{1}, \tau_{\delta} V)$.

The second term of \eqref{eq:multi_mukt} captures the conditional intensity component for counter-calls at MARU $k$ at time $t$. It is natural to assume that the intensity of counter-calls caused by an up-call at $t_i$ is higher in time near $t_i$ and higher in space near $m_i$ and decreases as we move forward in time and away in space.  Some motivation for this specification is provided by Figure~\ref{fig:ccbCor10}. We assume the $\alpha_{k}$'s are independent with a Gamma prior.

In this manuscript we consider the following four models.
\begin{enumerate} [(i)]
    \item NHPP: $\lambda_k(t \mid \mathcal{H}_t; \btheta) = e^{\beta_{0k} + \bx_k(t)^\top \boldsymbol{\beta}_k}$
    \item NHPP+GP: $\lambda_k(t \mid \mathcal{H}_t; \btheta) = e^{\beta_{0k} + \bx_k(t)^\top \boldsymbol{\beta}_k + \delta_k w(t)}$
    \item NHPP+CC: $\lambda_k(t \mid \mathcal{H}_t; \btheta) = e^{\beta_{0k} + \bx_k(t)^\top \boldsymbol{\beta}_k}+\sum_{i:t_i < t} \alpha_{m_i} e^{-\eta(t-t_i)} e^{-\phi d_{m_i,k}}$
    \item NHPP+GP+CC: $\lambda_k(t \mid \mathcal{H}_t; \btheta) = e^{\beta_{0k} + \bx_k(t)^\top \boldsymbol{\beta}_k + \delta_k w(t)}+\sum_{i:t_i < t} \alpha_{m_i} e^{-\eta(t-t_i)} e^{-\phi d_{m_i,k}}$
\end{enumerate}
Models (i)-(iii) are submodels of Model (iv). Models (i) and (ii) assume all up-calls are contact calls and that there are no counter-calls. Models (i) and (iii) are suitable when the covariate effects are sufficient to explain the rate of contact calls. The likelihood function \citep{White2021} associated with Model (iv) is given by
\begin{align}
    L(\btheta \mid \mathcal{T}) = \exp \left\{ - \sum_{k=1}^K \int_{0}^T \lambda_k(t \mid \mathcal{H}_{t}) dt \right\} \prod_{i=1}^n \lambda_{m_i}(t_i \mid \mathcal{H}_{t_i}).
    \label{eq:lik}
\end{align}
Model fitting details are found in the supplementary material.

\subsection{Novel inference}
\label{subsec:novinf}


The compensator for a point process provides the expected number of events under the intensity in the time window $(0,t]$.
The compensator for a process on $(0,T]$ with intensity $\lambda(t \mid \mathcal{H}_{t}; \btheta)$ is defined as $\Lambda(t \mid \mathcal{H}_{t}; \btheta) = \int_{0}^{t}\lambda(u \mid \mathcal{H}_{u}; \btheta) du$.  For any $t$, the compensator is a function of the data, i.e., it is a random variable.  Here we focus on inference using the compensator. We will employ the compensator for assessing model adequacy in Section~\ref{subsec:rtct}. If we evaluate the compensator at $t=T$, i.e., $\Lambda(T \mid \mathcal{H}_{T}; \btheta)$, we obtain the expected total number of events given the data.  If we separate the integral of the conditional intensity, we obtain the expected number of contact calls and counter-calls received, respectively, in the time window, $(0,T]$, given the data. Using an earthquake analogy, this separates the expected number of earthquakes and aftershocks. 

First, suppose that there is a single MARU. If we integrate the counter-call component, i.e., $\sum_{i:t_{i} \in \mathcal{T} } \int_{t_{i}}^{T} \alpha e^{-\eta(t-t_{i})}dt$, we obtain $ \alpha / \eta \sum_{i:t_{i} \in \mathcal{T}}[ 1-e^{-\eta(T-t_{i})} ]$. 
We can interpret the $i$th term in the sum as the contribution of the event at time $t_i$ to the expectation.  From the Bayesian perspective, given the data, using posterior samples of the model parameters, we can obtain the posterior distributions of these functions, i.e., the joint distribution of the expected number of contact calls and the expected number of counter-calls, their marginal distributions, and the distribution of the sum.

Consider an array of $K$ MARUs. The intensity of up-calls received by each MARU $k$ is specified by $\lambda_{k}(t \mid \mathcal{H}_t; \btheta) = \mu_{k}(t) + \sum_{i: t_{i}<t} \alpha_{m_{i}} e^{-\eta(t-t_{i})}e^{-\phi d_{m_{i}k}}$, which splits the conditional intensity into an intensity for contact calls received by MARU $k$ and an intensity for counter-calls received by MARU $k$. 
An up-call received at any MARU at any time prior to $t$ provides a contribution to the counter-call intensity for MARU $k$ at time $t$. This is the interpretation of $\sum_{i: t_{i}<t} \alpha_{m_{i}} e^{-\eta(t-t_{i})}e^{-\phi d_{m_{i}k}}$. 
The expected number of contact calls received by MARU $k$ over $(0, T]$ is $\int_0^T \mu_k(t) dt$. The expected number of counter-calls received at MARU $k$ over $(0, T]$ is $\int_{0}^{T}$ $\sum_{t_i < t}$ $\alpha_{m_i}$ $e^{-\eta(t-t_i)}$ $e^{-\phi d_{m_i, k}} dt = \sum_{i:t_i \in \mathcal{T}} \int_{t_{i}}^{T} \alpha_{m_i} e^{-\eta(t-t_i)} e^{-\phi d_{m_i, k}} dt = \sum_{i:t_i \in \mathcal{T}} (\alpha_{m_{i}}/\eta) [1- e^{-\eta(T-t_{i})}] e^{-\phi d_{m_i, k}}$. With multiple MARUs, if we sum the contact call component over $k$ we obtain the overall expected number of contact calls. If we sum the counter-call component over $k$ we obtain the overall expected number of counter-calls.  As above, we can obtain the joint posterior distribution of the expected number of contact calls and the expected number of counter-calls, their marginal distributions, and the posterior distribution of total expected number of up-calls. We can also obtain the expected number of counter-calls at MARU $k$ that were excited by up-calls at a specific MARU $\ell$ in the time window $(0, T]$. This is given by $\sum_{i:t_i \in \mathcal{T}} \mathbb{I}(m_i = \ell) (\alpha_{m_{i}}/\eta) [1- e^{-\eta(T-t_{i})}] e^{-\phi d_{m_i, k}}$ where $\mathbb{I}(\cdot)$ is the indicator function.

\section{Model assessment}
\label{sec:assess}

\subsection{RTCT and its implementation in a Bayesian framework}
\label{subsec:rtct}

We define the compensator associated with our model as $\Lambda(t \mid \mathcal{H}_t; \btheta) = \int_{0}^{t} \lambda(u \mid \mathcal{H}_{u}; \btheta) du$ as in Section~\ref{subsec:novinf}. By the random time change theorem (RTCT) \citep{Brown1988}, if $\btheta$ is ``true'' and $t^{*}_{i} = \Lambda(t_{i} \mid \mathcal{H}_{t_i}; \btheta)$ for $i=1,2,...,n$, then the set $\{t^{*}_{i}\}$ are a realization from a homogeneous Poisson process with rate 1 (HPP(1)) and the associated $d^{*}_{i} = t^{*}_{i} - t^{*}_{i-1}$ with $t^{*}_{0}=0$ are an i.i.d. sample from an exponential distribution with rate 1 (Exp(1)). 
For example, consider data $\mathcal{T}$ generated from an HPP$(\lambda)$. The associated waiting times $d_i = t_{i} - t_{i-1}$ with $t_0 = 0$ are independent and have an Exp$(\lambda)$ distribution. The compensator associated with HPP($\lambda$) is $t^{*}_i = \int_{0}^{t_i} \lambda du =  \lambda t_i$. Then the associated $d^{*}_i = t^{*}_{i} - t^{*}_{i-1} = \lambda(t_i - t_{i+1}) = \lambda d_i$ has an Exp(1) distribution.



The compensator $\Lambda(t \mid \mathcal{H}_{t}; \btheta)$, as a function of the data, is a random function. In the frequentist setting, there is a ``true'' $\btheta$ and only $\mathcal{T}$ is random. We have to estimate $\btheta$ and then use the estimated compensator $\Lambda(t \mid \mathcal{H}_{t}; \hat{\btheta})$.  This yields $\{ d^{*}_{i}(\mathcal{T}; \hat{\btheta}) \}$.  The RTCT does not formally apply to $\{ d^{*}_{i}(\mathcal{T}; \hat{\btheta}) \}$ but the argument is that, if $\hat{\btheta}$ is close to the true $\btheta$, the result holds approximately.

We consider the RTCT in a fully Bayesian modeling setting.  From a frequentist perspective the compensator is a random variable as a function of the data.  From the Bayesian perspective it is a random variable as a function of the parameters given the data and we examine it through posterior inference. 
Consider the simple HPP($\lambda$) case. If we draw $\lambda_{b} \sim p(\cdot)$ a priori and $\mathcal{T}_{b} \sim$ HPP$(\lambda_{b})$, then the set $\{d_{b,i}^{*}\}$ are an i.i.d. sample from Exp(1). This is also true if we generate $\mathcal{T} \sim$ HPP$(\lambda)$ and define $\lambda_{b} \sim \pi(\cdot \mid \mathcal{T})$ as the posterior distribution. If the model is correctly specified, the posterior sample $\{d_{b,i}^{*}\}$ are a realization from Exp(1). In other words, if the posterior sample $\{d_{b,i}^{*}\}$ do not look like Exp(1) samples, then the model is not well specified. The same argument applies to a general point process model.

For the multivariate Hawkes process model, model assessment using RTCT only requires evaluating $d_{b,i}^{*} = t_{b,i}^{*}- t_{b,i-1}^{*} = \int_{t_{i-1}}^{t_{i}} \lambda(u \mid \mathcal{H}_{u}; \btheta_b) du$ for $i = 1,\dots,n$ and for each posterior sample $\btheta_b$. We can obtain the estimated posterior mean $\hat{d}_{i}^{*}$ of $d_{i}^{*}$ with associated uncertainty and examine if it resembles a sample from Exp(1).

\subsection{Model adequacy and model comparison}
\label{subsec:qqplotDIC}

To examine model adequacy, we use the Q-Q plot for $\{d_{i}^{*}\}$ against an Exp(1) distribution. Specifically, we obtain the order statistics $\{\hat{d}_{(i)}^{*}\}$ of the estimated posterior means $\{\hat{d}_{i}^{*}\}$ and plot them against the theoretical quantiles for an Exp(1), i.e., $\log\frac{n}{n- i - 0.5}$. The Q-Q plot can be overlaid with the associated 95\% credible band. We look for fidelity to a $45^{o}$ reference line as an assessment of model adequacy.

In this manuscript we consider four different models:  (i) NHPP, (ii) NHPP+GP, (iii) NHPP+CC, (iv) NHPP+GP+CC. Visual comparison of the Q-Q plots, with associated uncertainties provides informal model comparison. Using the metric $\frac{1}{n}\sum_{i} \big(\hat{d}_{(i)}^{*} -\log\frac{n}{n - i - 0.5}\big)^{2}$, which quantifies the mean squared difference between sample and theoretical quantiles, we can assign a measure of model performance associated with the Q-Q plot.

For model comparison we consider DIC = $\widehat{-2 \log L} + p_{D}$. The first term is the estimated posterior mean of $-2 \log L(\btheta \mid \mathcal{T})$ and rewards model fit. The second term $p_{D} = \widehat{-2 \log L} + \log L (\hat{\btheta} \mid \mathcal{T})$, where $\hat{\btheta}$ is the estimated posterior mean of $\btheta$, is the estimated effective number of parameters and penalizes model complexity. The DIC is easily evaluated using posterior samples in the Bayesian framework.

\section{Simulation study}
\label{sec:sim}

The aims of this simulation experiment are to evaluate the capability of fitting models to identify generating models and to demonstrate the ability to distinguish between the GP and CC components of the intensity.

We simulate a dataset from each of the four models (i) NHPP, (ii) NHPP+GP, (iii) NHPP+CC, and (iv) NHPP+GP+CC. We use the actual locations of $K = 10$ MARUs in CCB and consider 5 consecutive days for the observed time window. For each MARU $k$, we use $\mu_k(t)$ to generate a set of contact call times from an NHPP employing the method of \citet{Lewis1979}. This requires sampling from an HPP with intensity of $\max_{t \in (0, T]}\{\mu_k(t)\}$ and using rejection sampling to obtain the resulting event sequence. Aggregating across the MARUs, we have a set of contact call times. Each up-call time $t_i$ has a mark $m_i$ indicating the MARU that received the up-call. We can order them as usual to create $\{(t_i,m_i)\}$. 

\begin{figure}[!b] 
\begin{center}
\includegraphics[width=0.8\textwidth]{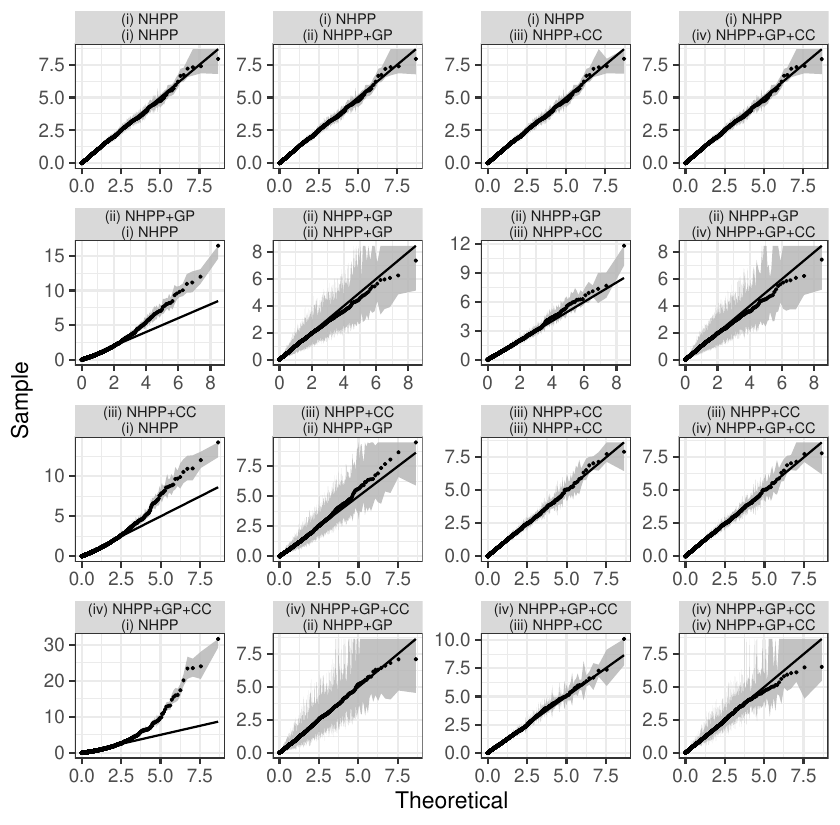}
\end{center}
\caption{Q-Q plots for $\hat{d}^{\ast}_i$'s against an Exp(1) distribution. Gray shades represent 95\% credible bands. Top and bottom labels denote generating and fitting models, respectively. \label{fig:msimQQband}}
\end{figure}

For Models (iii) and (iv), we generate counter-calls associated with each of the contact calls. For each $\{(t_i,m_i)\}$, we generate the offspring at MARU $k$ using the counter-call intensity associated with MARU $k$, i.e., $\alpha_{m_i} e^{-\eta(t-t_i)} e^{-\phi d_{m_i,k}}$. To do this, we again simulate from an NHPP and use the Lewis and Shedler method to generate events on $(t_i, T]$. Note that there is no generation of marks; these events all have mark $k$. This process is repeated for each $k$ and all $\{(t_i,m_i)\}$ starting from $\{(t_1,m_1)\}$. The entire resulting collection of events becomes a realization of a multivariate Hawkes process. Importantly, this simulation method provides a generating model that agrees with the fitting model described above.

For all but $\beta_{0k}$, the true values of the model parameters are set to the estimated posterior means from Model (iv) fitted to the multi-channel dataset (as discussed in Section~\ref{subsec:mreal}). We choose $\beta_{0k}$ such that the total number of up-calls is approximately equal to what is observed in the real data.

We fit Models (i) to (iv) to each of the four simulated datasets using Markov chain Monte Carlo (MCMC) run for 100,000 iterations. The first 10,000 posterior samples are discarded as burn-in and the remaining 90,000 are used for posterior inference. We examined trace plots of chains for convergence and no issues were detected.

\setlength{\tabcolsep}{3pt} 
\begin{table}[!tb]
\caption{DIC from Models (i) to (iv) fitted to data generated from each of the models. All entries are of order $10^{4}$. Bolds denote the fitting models with the smallest DIC.
\label{tab:msimDIC}}
\small
\begin{center}
\begin{tabular}{lrrrr}
  \toprule
Generating $\backslash$ Fitting & (i) NHPP & (ii) NHPP+GP & (iii) NHPP+CC & (iv) NHPP+GP+CC \\ 
  \midrule
(i) NHPP & \textbf{2.24} & \textbf{2.24} & \textbf{2.24} & \textbf{2.24} \\ 
  (ii) NHPP+GP & 1.93 & \textbf{1.84} & 1.89 & \textbf{1.84} \\ 
  (iii) NHPP+CC & 2.15 & 2.09 & \textbf{2.03} & \textbf{2.03} \\ 
  (iv) NHPP+GP+CC & 2.18 & 2.02 & 1.96 & \textbf{1.94} \\ 
   \bottomrule
\end{tabular}
\end{center}
\end{table}

We use RTCT and DIC to assess how well fitting models can distinguish generating models. Figure~\ref{fig:msimQQband} displays the posterior mean empirical Q-Q plots for $d^{\ast}_i$'s against an Exp(1) distribution for each combination of generating and fitting models. The points of (i) NHPP follow the 45$^{\circ}$ reference line well for data generated from (i) NHPP, however they significantly deviate from the reference line for data from the other models. This implies that Model (i) only fits data generated under Model (i).
Models (ii) to (iv) seem to fit every dataset fairly well. Their points generally fall on the reference line. This may imply that the GP and CC components are flexible enough to account for variability from other sources. The 95\% credible band (CB) of Models (ii) NHPP+GP and (iv) NHPP+GP+CC are wider than the other models, attributing to the extra uncertainty from the GP. Table~\ref{tab:msimDIC} shows that DIC correctly selects the true model and model(s) that have the true model as a submodel. 


\begin{figure}[!tb] 
\begin{center}
\includegraphics[width=\textwidth]{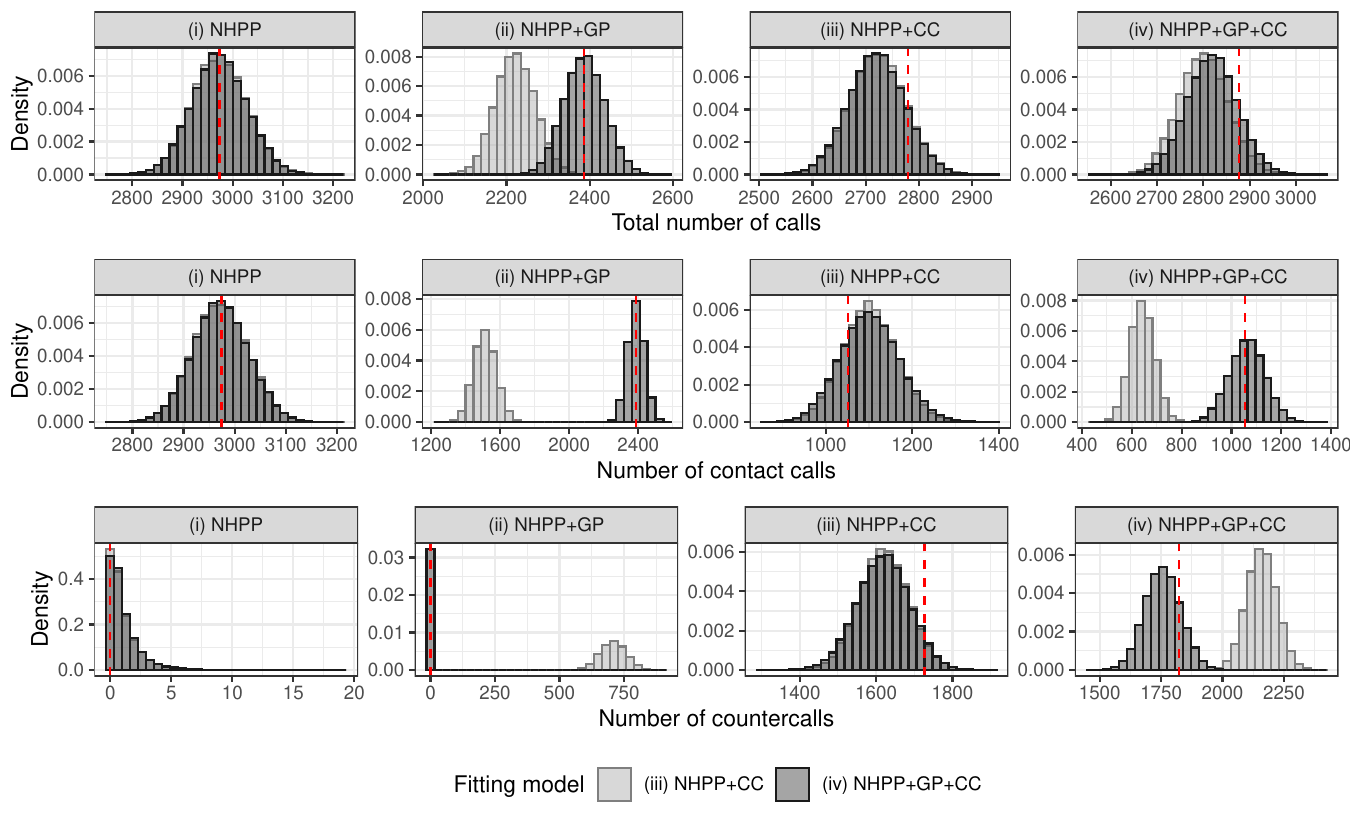}
\end{center}
\caption{Posterior distributions of the expected 
total number of up-calls (top), the expected number of contact calls (middle), and the expected number of counter-calls (bottom) from (iii) NHPP+CC (light gray) and (iv) NHPP+GP+CC (dark gray). Facet labels denote generating models. Red dashed vertical lines are the observed values. \label{fig:msimNum34}}
\end{figure}

We estimate the expected total number of up-calls, expected number of contact calls, and expected number of counter-calls for each scenario. Each model recovers the true expected total number of up-calls well for every dataset. When there are counter-calls in the generated data, Models (i) and (ii), which have no CC component, overestimate the expected number of contact calls to recover the expected total number of up-calls. Figure~\ref{fig:msimNum34} displays the empirical posterior distributions for the expectations from Models (iii) NHPP+CC and (iv) NHPP+GP+CC. Model (iii) recovers the simulated truth well for datasets generated from itself and its submodel, i.e., Model (i) and (iii). When there is a GP in the generative model, Model (iii) considerably underestimates the expected number of contact calls and overestimates the expected number of counter-calls. This implies that the CC component of Model (iii) inappropriately accounts for variability from the GP component of the intensity. Model (iv) recovers the truth fairly well for every dataset. This implies that the GP component is well distinguished from the CC component of the intensity. See the supplementary material for additional simulation results.

\section{Analysis of North Atlantic right whale up-call data}
\label{sec:real}


\subsection{Single-channel data}
\label{subsec:sreal}

First we analyze NARW up-calls observed by a single recorder (Section~\ref{subsec:sdata}). We set the number of MARUs $K = 1$ and fit Models (i)-(iv) described in Section~\ref{sec:models}.

\begin{figure}[!b] 
\begin{center}
\includegraphics[width = 0.85\textwidth]{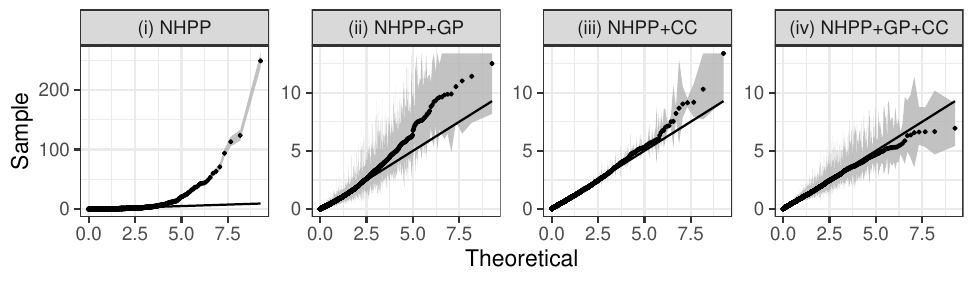}
\end{center}
\caption{Q-Q plots for $\hat{d}^{\ast}_i$'s against an Exp(1) distribution. Gray shades represent 95\% credible bands. Facet labels denote models fitted to the single-channel dataset.\label{fig:noppQQband}}
\end{figure}

We assess model adequacy using the RTCT. Figure~\ref{fig:noppQQband} displays the posterior mean empirical Q-Q plots for $\{d^{\ast}_i, i = 1,\dots,n\}$ against an Exp(1) distribution.
For Model (i) the points significantly deviate from the 45$^{\circ}$ reference line, which is not captured by the 95\% credible band (CB). This implies that the NHPP is not adequate for this dataset. 
The 95\% CBs of Models (ii) and (iv) are wider than the other models, attributable to the extra uncertainty from the Gaussian process, and do include the reference line. The points of Models (iii) and (iv) follow the line best although they slightly curve away near the upper tail. As for the mean squared difference (MSD) between sample and theoretical quantiles, Model (iv) provides the smallest MSD = 0.006, followed by Model (iii) with the second smallest MSD = 0.012, Model (ii) with the third smallest MSD = 0.08, and Model (i) with the largest MSD = 24.02.

\begin{table}[!tb] 
\caption{DIC from Models (i) to (iv) fitted to the single-channel dataset. We compare DIC up to three significant digits. 
Bold represents the model with the smallest DIC.\label{tab:srealDIC}}
\small
\begin{center}
\begin{tabular}{l rrr}
  \toprule
  Model & $\widehat{-2 \log L}$ & (95\% HPD)  & DIC\\ 
  \midrule
  (i) NHPP & 14159 & (14152, 14166) & 14167 \\ 
  (ii) NHPP+GP &  7646 & (7588, 7695) &  7874 \\ 
  (iii) NHPP+SE &  7834 & (7826, 7843) &  7843 \\ 
  (iv) NHPP+GP+SE &  7452 & (7384, 7530) & \textbf{7558} \\
  \bottomrule
\end{tabular}
\end{center}
\end{table}

We use DIC for model comparison. Table~\ref{tab:srealDIC} shows that Model (iv) has a significantly small value of $-2 \log L$ compared to Models (i) and (iii). This implies that Model (iv) fits the data significantly better than Models (i) and (iii). Model (ii) fits the data similarly well to Model (iv) but has more than twice as large of a model complexity penalty as Model (iv). This may indicate that the GP component of Model (ii) inappropriately accounts for variability from excitement. DIC provides consistent results with MSD. Model (iv) has the smallest DIC, followed by Model (iii) with the second smallest DIC, Model (ii) with the third smallest DIC, and Model (i) with the largest DIC. Collectively, these metrics support our choice of Model (iv) for this dataset and for which we present further model inference.

\begin{figure}[!b]
\begin{center}
\includegraphics[width = 0.35\textwidth]{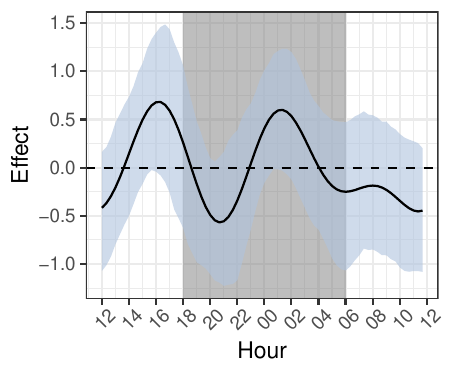}
\end{center}
\caption{Estimated daily harmonic effect over time. Blue shade indicates 95\% credible bands. Gray shade indicates 6 p.m. to 6 a.m. in local time.\label{fig:noppXB}}
\end{figure}

The estimated posterior mean of the coefficient $\beta_1$ of the ambient noise variable is -0.49 with 95\% HPD interval of (-0.71, -0.28). This significant effect implies that NARWs are less likely to make up-calls during periods with high levels of ambient noise. The observed responses
of NARWs to increased levels of noise include changes in sound level \citep{parksIndividualRightWhales2010, palmerAccountingLombardEffect2022}, frequency content  \citep{urazghildiievStatisticalAnalysisNorth2014} and up-calling rate (CWC pers obs.). Here we quantify the reduction in up-calling behavior as background noise levels increases, a response that  is consistent with previous work on  NARWs in CCB   \citep{urazghildiievStatisticalAnalysisNorth2014} and bowhead whales (\textit{Balaena mysticetus}) in the Arctic \citep{blackwellEffectsAirgunSounds2013}.
Figure~\ref{fig:noppXB} shows the resulting estimated overall daily effect as captured by the harmonics over time. Shown is the posterior estimate of $\hat{\beta}_{2} \sin(\frac{2 \pi t}{8 \times 60}) + \hat{\beta}_{3} \cos(\frac{2 \pi t}{8 \times 60}) + \hat{\beta}_{4} \sin(\frac{2 \pi t}{12 \times 60}) + \hat{\beta}_{5} \cos(\frac{2 \pi t}{12 \times 60}) + \hat{\beta}_{6} \sin(\frac{2 \pi t}{24 \times 60}) + \hat{\beta}_{7} \cos(\frac{2 \pi t }{24 \times 60})$ over the 24-hour window, which better depicts the diel behaviors than any of the individual coefficient estimates separately. 
Contact calls tend to increase during the twilight hours between dusk and sunset, as well as around midnight, which has been well documented in NARW across a broad range of habitats \citep{matthews2001vocalisation,matthewsOverviewNorthAtlantic2021}.

From this analysis of NARW up-call times of occurrence, we find strong evidence of \textit{excitement}. The estimated posterior mean of the \textit{excitement} parameter $\alpha$ is 0.34 with 95\% HPD interval of (0.31, 0.38), which is much greater than 0. The estimated posterior mean of the decay parameter $\eta$ is 0.51 with 95\% HPD interval of (0.44, 0.58). This implies that the expected median response time for NARW up-calls is approximately $\frac{\log(2)}{0.51} = 1.36$ minutes, consistent with direct measurements of up-calls from animals tagged with DTAGs in other NARW habitats \citep{parksIndividualRightWhales2010}. During the study period, the estimated posterior mean for the expected total number of up-calls is 5301 with 95\% HPD interval of (5162, 5448), which contains the observed total number of up-calls of 5302. The estimated posterior means of the expected number of contact calls and counter-calls are 1735 with 95\% HPD interval of (1364, 2066) and 3567 with 95\% HPD interval of (3223, 3947), respectively. We observe approximately twice as many counter-calls as contact calls on average.

\subsection{Multi-channel CCB data}
\label{subsec:mreal}

To study spatio-temporal patterns of NARW up-calls across CCB, we apply Models (i)-(iv) described in Section~\ref{sec:models} to the sequence of up-call times received by the array of ten MARUs within the region (Section~\ref{subsec:mdata}).


\begin{figure}[!b]
\begin{center}
\includegraphics[width = 0.85\textwidth]{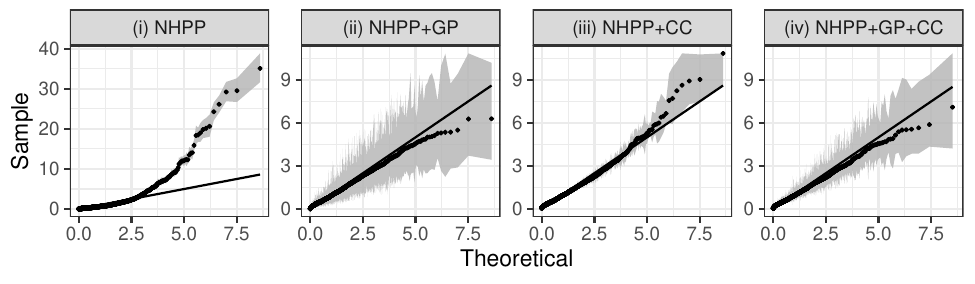}
\end{center}
\caption{Q-Q plots for $\hat{d}^{\ast}_i$'s against an Exp(1) distribution. Gray shades represent 95\% credible bands. Facet labels denote models fitted to the multi-channel dataset.\label{fig:mrealQQband}}
\end{figure}

Model adequacy for the multi-channel data is assessed using the RTCT.  Figure~\ref{fig:mrealQQband} shows that Model (i) is again inadequate for these NARW up-call data, as evident by the points and 95\% credible band (CB) that significantly deviate from the reference line. The points of Models (ii) to (iv) follow the line better, but deviate slightly near the upper tail. As with the results for the single-channel data, the 95\% CBs of Models (ii) and (iv) are wide due to the extra uncertainty from the Gaussian process; only these two models completely include the reference line. As for the mean squared difference (MSD) between sample and theoretical quantiles, Models (ii) to (iv) provide the smallest MSD = 0.02 and Model (i) provides the largest MSD = 1.87.

\begin{table}[!tb]
\caption{DIC from Models (i) to (iv) fitted to the multi-channel dataset. We compare DIC up to three significant digits. Bold represents the model with the smallest DIC.\label{tab:mrealDIC}}
\small
\begin{center}
\begin{tabular}{l rrr}
  \toprule
  Model & $\widehat{-2 \log L}$ & (95\% HPD) & DIC\\ 
  \midrule
  (i) NHPP & 24483 & (24460, 24508) & 24558 \\
  (ii) NHPP+GP & 22501 & (22435, 22564) & 22842 \\ 
  iii) NHPP+SE & 22130 & (22105, 22155) & 22196 \\
  (iv) NHPP+GP+SE & 21712 & (21635, 21792) & \textbf{21908} \\
  \bottomrule
\end{tabular}
\end{center}
\end{table}

Table~\ref{tab:mrealDIC} reports the DIC values for model comparison. Notably, Model (iv) has a significantly smaller value of $-2 \log L$ than the other models. Model (iv) also achieves the smallest DIC, indicating this is the preferred model for these data.
We again select Model (iv), that with the GP and CC, as our final model for this dataset and present the associated results. 

\begin{figure}[!t] 
\begin{center}
\includegraphics[width = \textwidth]{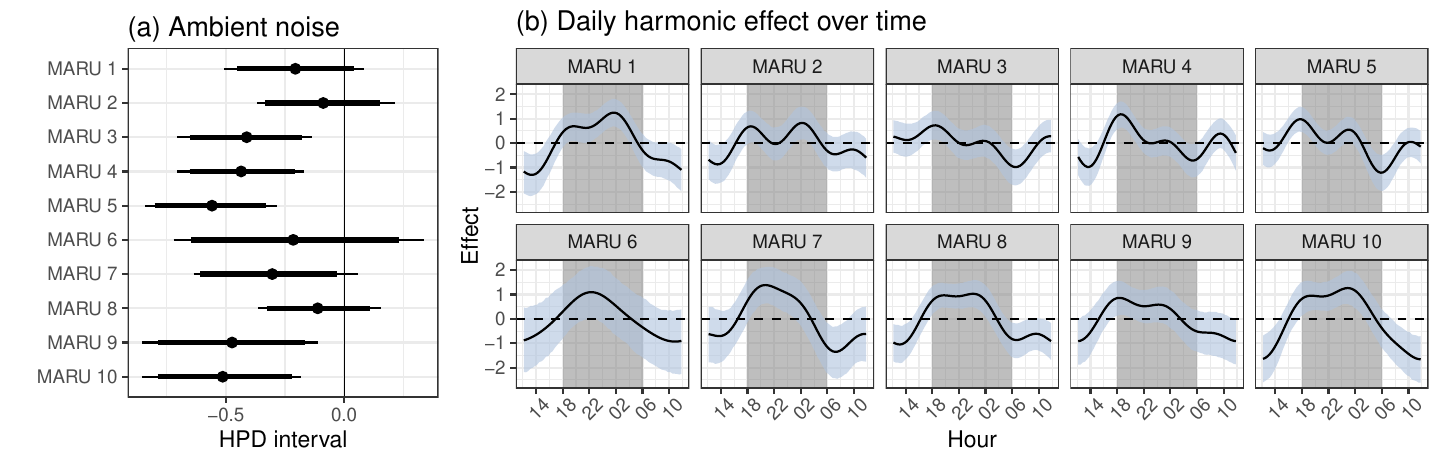}
\end{center}
\caption{(a) Estimated posterior means (shaded dots), 95\% HPD intervals (thin horizontal bars), and 90\% HPD intervals (thick horizontal bars) for the coefficient of the ambient noise per MARU. (b) Estimated diel harmonic effects over time for each MARU. Blue shades are 95\% credible bands. Gray shades indicate 6 p.m. to 6 a.m. in local time. \label{fig:mrealCInXB}}
\end{figure}

Figure~\ref{fig:mrealCInXB} (a) displays the estimated posterior mean and HPD intervals of the noise coefficient $\beta_{1k}$ for each MARU $k$. The posterior mean estimate is negative for each MARU, and the coefficient is significantly negative as evident by the 95\% HPD for MARUs located in the bottom edge of the array, i.e., MARUs 3, 4, 5, 9, and 10. The 95\% HPD intervals of MARU 1 and 7 include zero but have upper bounds very close to zero. MARU 6 has a very wide HPD interval, which we attribute to the small sample size for this MARU; the expected number of unique up-calls at MARU-6 is estimated to be only 26. Figure~\ref{fig:mrealCInXB} (b) shows the estimated overall daily effect as captured by the harmonics over the 24-hour window for each MARU. In general, contact calls tend to increase during the twilight hours between dusk and sunset. We observe a significant increase in up-calling during the night for MARUs 1, 7, 8, and 10. Since MARU 10 is close to a shipping lane, it's possible that up-calling increases at nighttime with lower shipping. NARW have been known to up-call more at night \citep{matthews2001vocalisation, matthewsOverviewNorthAtlantic2021}; however because we lack visual observations during this period, we cannot say for sure if this is in response to non-foraging aggregations or distribution changes within and across the bay.

\begin{figure}[!b] 
\begin{center}
\includegraphics[width = 0.9\textwidth]{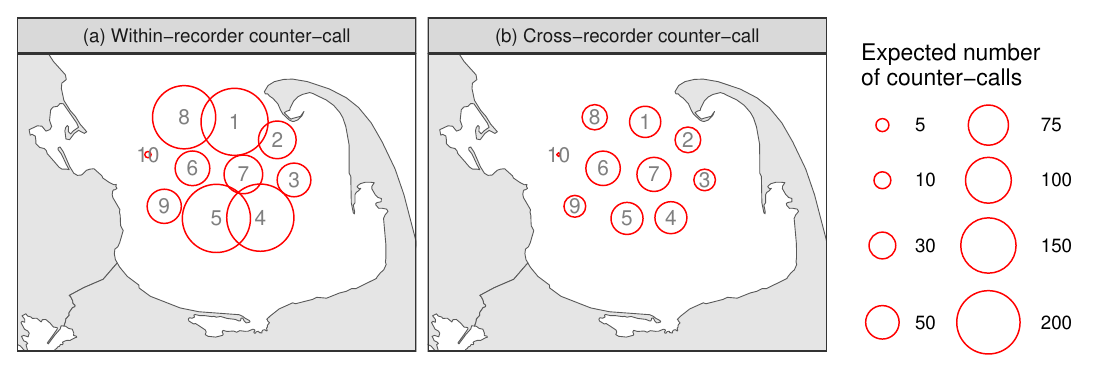}
\end{center}
\caption{(a) Estimated posterior mean for the expected number of within-recorder counter-calls received at each MARU. (b) Estimated posterior mean for the expected number of cross-recorder counter-calls received at each MARU. The number of counter-calls is shown proportionally by the size of the circle.
\label{fig:mrealExcitement}}
\end{figure}

Except for MARU 10, we find strong evidence of excitement in NARW up-calls; the 95\% HPD interval of $\alpha_k$ is greater than zero for $k = 1,\dots,9$. MARU 10 is located in a region with the highest shipping traffic. This might prevent whales from hearing and responding to up-calls and disturb communication between whales. The estimated posterior mean of the temporal decay parameter $\eta$ is 0.151 with 95\% HPD interval of (0.150, 0.154). This implies that the expected median response time for NARW up-calls is approximately $\frac{\log(2)}{0.151} = 4.59$ minutes. 
The estimated posterior mean of the spatial decay parameter $\phi$ is 0.32 with 95\% HPD interval of (0.29, 0.35). The shortest distance between MARUs is 7.1 km. This implies that the probability of a counter-call being received at a MARU at least 7.1km away is approximately $P(\text{Exp}(0.32) > 7.1) = 0.10$.

During the study period, the estimated posterior mean for the expected total number of up-calls is 2716 with 95\% HPD interval of (2613, 2820) which contains the observed total number of up-calls of 2750. The model also recovers the observed number of up-calls recorded by each MARU within the 95\% HPD intervals. Figure~\ref{fig:mrealExcitement} (a) displays the estimated posterior mean for the expected number of within-recorder counter-calls, i.e., counter-calls at MARU $k$ that were excited by calls at the MARU. Figure~\ref{fig:mrealExcitement} (b) shows the estimated posterior mean for the expected number of cross-recorder counter-calls, i.e., counter-calls at MARU $k$ that were excited by up-calls at MARU $\ell$ for $\ell \neq k$. We see that MARU 10 received a relatively small number of within- and cross-recorder counter-calls. MARU 10 is closest to the path of vessels that transit from the Cape Cod Canal to Boston Harbour. It is the loudest MARU (Figure~\ref{fig:noiseCCB}), which may indicate that underwater noise caused by shipping can inhibit whales from being able to communicate with each other. Elevated ambient noise can also impact our ability to detect up-calls \citep{palmerAccountingLombardEffect2022}. MARUs 1, 4, 5, and 8 received a substantial number of within-recorder counter-calls compared to the others. In contrast to MARU 10, MARU 5 was one of the quietest MARUs (Figure~\ref{fig:noiseCCB}). MARUs 1 and 8 are at the top of the array (Figure~\ref{fig:ccbHP}), and thus are likely hearing up-calls that are made outside of CCB as well. CCB is a relatively small area, and sound can be heard across much of the Bay; the average swim speed of NARW is approximately 3.7 km per hour, and the shortest distance between MARUs is 7.1 km. Thus, the excitement across recorders likely indicates communication between different whales, which is consistent with the role of the contact call \citep{clark1980sound,clark1982}. 

\begin{figure}[!tb] 
\begin{center}
\includegraphics[width = \textwidth]{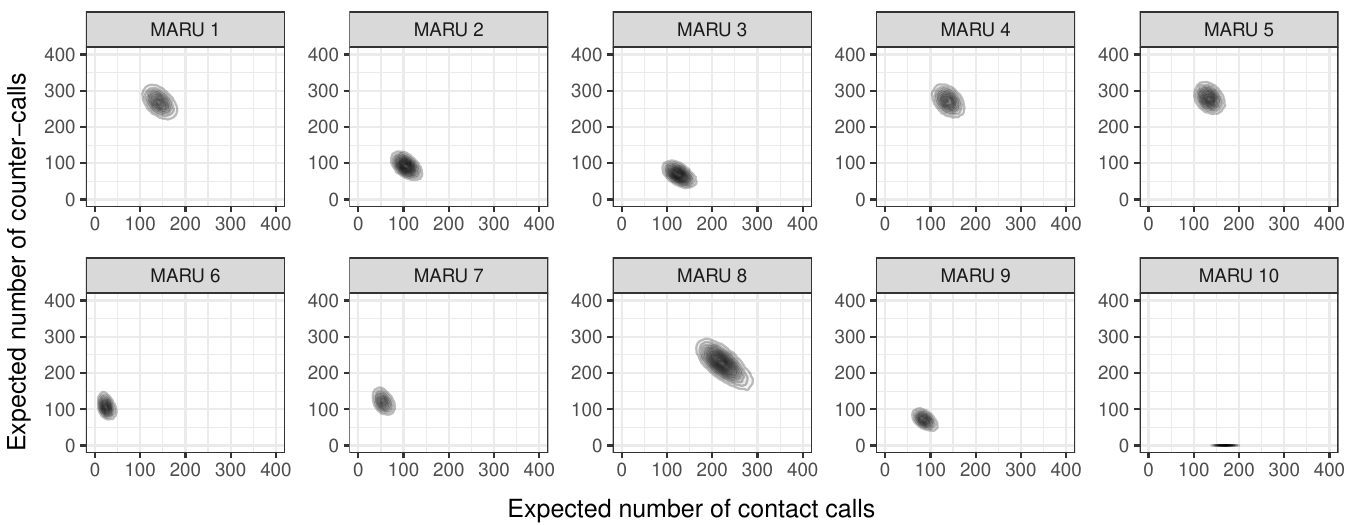}
\end{center}
\caption{Joint posterior distribution of the expected number contact calls and counter-calls received at each MARU.
\label{fig:mrealNumJoint}}
\end{figure}

Figure~\ref{fig:mrealNumJoint} shows the joint posterior distribution of the expected number contact calls and counter-calls received at each MARU. The majority of MARUs received a similar or greater number of counter-calls compared to contact calls. Uniquely, MARU 3 received twice as many contact calls as counter-calls and MARU 10 received mostly contact calls and very few counter-calls. More detailed evidence of this behavior is presented in the supplementary material. Figure S5 displays the posterior mean estimates of the background intensity and the counter-call intensity. The panel for MARU 10 further depicts the sparsity of expected counter-calls there, reflecting the estimate of $\alpha_{10}$ which is essentially $0$. 

Figure S6 of the Supplementary Material aggregates the two intensity components for each MARU and overlays, with rug tassels, the observed event sequence (Figure~\ref{fig:noiseCCB} (b)).  It offers informal/visual evidence of good model performance for the observed data.

\section{Summary and future work}
\label{sec:summary}

We have developed a multivariate Hawkes process model framework to address the current knowledge gap on the communication behavior between NARWs over space and time. The class of models are informed from sequences of event times for NARW up-calls collected at a small network of recorders in Cape Cod Bay, MA. This is the first network level investigation of such up-calls, enabling different inference than what can be obtained from putting acoustic tags on individual animals. As mentioned in the introduction, acoustic data are often collected from three sources: 1) tags deployed on animals; 2) arrays supplemented with visual tracking data; or 3) towed or fixed arrays. Each provides a different snapshot of acoustic behavior. We believe our inferential approach provides an excellent and needed way to examine inter-whale communication across an array.

Our models consider the incidence of contact calls as well as counter-calls using multi-channel excitement models.  We specify conditional intensities for each channel in the network which enable contact calls to be described using fixed and random effects and counter-calls using a mutually exciting conditional component. We find that the model incorporating a nonhomegenous Poisson process, a Gaussian process, and counter-calling to be preferred over simpler models. As such, the richer models allow us to uncover subtler patterns of acoustic behavior among NARWs. 

One important limitation of our work is that while we infer about whale communication, what we are modeling are up-calls received at the set of recorders.  That is, we consider the behavior of received up-calls to be a proxy for the behavior of individual NARW up-calling. Specifically, we are modeling contact calling and counter-calling in the context of the event sequence of received up-calls for the network of recorders. Given the rapid decay specified in exciting behavior and the spatial locations of the recorders, if self or mutual excitement occurs across recorders, we are imagining that this is most plausibly the result of communication between different whales. Another limitation is that we have developed our dataset to ensure that a counter-call is associated with one and only one recorder and we have eliminated delays in up-calls, i.e., the time an up-call was made compared with the time it was received at a recorder. With manual annotation of up-calls \citep{palmerAccountingLombardEffect2022}, it may be able to identify the time and location of an up-call, which would likely enrich our understanding of the behavior.  Lastly, we do not account for acoustic detection \citep{palmerAccountingLombardEffect2022}, which may mean we are missing up-calls from whales that are either farther away from a MARU, or made during periods of elevated ambient noise. Here we worked only with up-calls that were sufficient in strength and number to be associated and localized. All of these issues present opportunities for future work.

Because NARW are not obligate callers \citep{parksSoundProductionBehavior2011}, incorporating auxiliary information about their observed behavior from aerial or vessel-based surveys may enable better understanding of state-specific calling rates \citep{matthewsOverviewNorthAtlantic2021}. Methods for estimating absolute abundance using up-call data have assumed that individual up-calling rates are known \citep{matthewsOverviewNorthAtlantic2021,schliep2023assessing}. Our work can provide a better understanding of the variation in NARW up-calling behavior across time and space. This may help better estimate whales' abundance and spatiotemporal distribution when employed with a combination of visual and acoustic observations.

\section*{Acknowledgements}
We thank the participants who attended a workshop immediately following the 2023 Annual Right Whale Consortium Meeting in Halifax, Nova Scotia for ideas and critiques. We thank Doug Gillespie for help with PAMGuard. This work was supported by the US Office of Naval Research Award N000141712817. Additional funding was provided by NOAA Fisheries under award NA20NMF0080246, and SERDP award RC20-1097. 


\bigskip
\begin{center}
{\large\bf SUPPLEMENTARY MATERIAL}
\end{center}

\begin{description}

\item[Supplementary material:] Model fitting details, tables and figures.




\end{description}



\bibliographystyle{apalike}
\bibliography{refs}

\begin{thebibliography}{}

\bibitem[Au and Hastings, 2008]{au2008principles}
Au, W.~W. and Hastings, M.~C. (2008).
\newblock {\em Principles of marine bioacoustics}, volume 510.
\newblock Springer.

\bibitem[Bass and Clark, 2003]{bass2003physical}
Bass, A.~H. and Clark, C.~W. (2003).
\newblock The physical acoustics of underwater sound communication.
\newblock In {\em Acoustic communication}, pages 15--64. Springer.

\bibitem[Baumgartner and Mussoline,
  2011]{baumgartnerGeneralizedBaleenWhale2011}
Baumgartner, M.~F. and Mussoline, S.~E. (2011).
\newblock A generalized baleen whale call detection and classification system.
\newblock {\em J. Acoust. Soc. Am.}, 129(5):2889--2902.

\bibitem[Blackwell et~al., 2013]{blackwellEffectsAirgunSounds2013}
Blackwell, S.~B., Nations, C.~S., McDonald, T.~L., Greene~Jr., C.~R., Thode,
  A.~M., Guerra, M., and Michael~Macrander, A. (2013).
\newblock Effects of airgun sounds on bowhead whale calling rates in the
  {{Alaskan Beaufort Sea}}.
\newblock {\em Marine Mammal Science}, 29(4):E342--E365.

\bibitem[Bradbury and Vehrencamp, 2011]{bradbury1998principles}
Bradbury, J.~W. and Vehrencamp, S.~L. (2011).
\newblock {\em Principles of animal communication, Second Edition}.
\newblock Sinauer Associates Sunderland, MA.

\bibitem[Brown and Nair, 1988]{Brown1988}
Brown, T.~C. and Nair, M.~G. (1988).
\newblock A simple proof of the multivariate random time change theorem for
  point processes.
\newblock {\em Journal of Applied Probability}, 25(1):210--214.

\bibitem[Calupca et~al., 2000]{calupca2000}
Calupca, T.~A., Fristrup, K.~M., and Clark, C.~W. (2000).
\newblock A compact digital recording system for autonomous bioacoustic
  monitoring.
\newblock {\em The Journal of the Acoustical Society of America},
  108(Supplement):2582--2582.

\bibitem[Clark, 1982]{clark1982}
Clark, C.~W. (1982).
\newblock The acoustic repertoire of the {{Southern}} right whale, a
  quantitative analysis.
\newblock {\em Anim. Behav.}, 30(4):1060--1071.

\bibitem[Clark et~al., 2010]{clarkVisualAcousticSurveys2010a}
Clark, C.~W., Brown, M.~W., and Corkeron, P. (2010).
\newblock Visual and acoustic surveys for {{North Atlantic}} right whales,
  {{Eubalaena}} glacialis, in {{Cape Cod Bay}}, {{Massachusetts}},
  2001{\textendash}2005: {{Management}} implications.
\newblock {\em Marine Mammal Science}, 26(4):837--854.

\bibitem[Clark and Clark, 1980]{clark1980sound}
Clark, C.~W. and Clark, J.~M. (1980).
\newblock Sound playback experiments with {Southern} right whales
  (\textit{Eubalaena australis}).
\newblock {\em Science}, 207(4431):663--665.

\bibitem[Clark et~al., 1986]{clark1986preliminary}
Clark, {\relax CW}., Ellison, {\relax WT}., and Beeman, K. (1986).
\newblock A preliminary account of the acoustic study conducted during the 1985
  spring bowhead whale, {{Balaena}} mysticetus, migration off {{Point Barrow}},
  {{Alaska}}.
\newblock {\em Report of the International Whaling Commission}, 36:311--316.

\bibitem[Cusano et~al., 2019]{cusanoImplementingConservationMeasures2019}
Cusano, D.~A., Conger, L.~A., Van~Parijs, S.~M., and Parks, S.~E. (2019).
\newblock Implementing conservation measures for the {{North Atlantic}} right
  whale: Considering the behavioral ontogeny of mother-calf pairs.
\newblock {\em Anim. Conserv.}, 22(3):228--237.

\bibitem[Davis et~al., 2017]{davisLongtermPassiveAcoustic2017}
Davis, G.~E., Baumgartner, M.~F., Bonnell, J.~M., Bell, J., Berchok, C.,
  Bort~Thornton, J., Brault, S., Buchanan, G., Charif, R.~A., Cholewiak, D.,
  Clark, C.~W., Corkeron, P., Delarue, J., Dudzinski, K., Hatch, L.,
  Hildebrand, J., Hodge, L., Klinck, H., Kraus, S., Martin, B., Mellinger,
  D.~K., {Moors-Murphy}, H., Nieukirk, S., Nowacek, D.~P., Parks, S., Read,
  A.~J., Rice, A.~N., Risch, D., {\v S}irovi{\'c}, A., Soldevilla, M.,
  Stafford, K., Stanistreet, J.~E., Summers, E., Todd, S., Warde, A., and
  Van~Parijs, S.~M. (2017).
\newblock Long-term passive acoustic recordings track the changing distribution
  of {{North Atlantic}} right whales ({{Eubalaena}} glacialis) from 2004 to
  2014.
\newblock {\em Sci. Rep.}, 7(1):13460.

\bibitem[Davis et~al., 2023]{davisUpcallingBehaviourPatterns2023}
Davis, G.~E., Tennant, S.~C., and Van~Parijs, S.~M. (2023).
\newblock Upcalling behaviour and patterns in {{North Atlantic}} right whales,
  implications for monitoring protocols during wind energy development.
\newblock {\em ICES Journal of Marine Science}, page fsad174.

\bibitem[{DCLDE 2013}, 2023]{dclde2023}
{DCLDE 2013} (2023).
\newblock {The North Atlantic right whale annotations from the detection
  classification localization and density estimate conference in 2013}.
\newblock {\em NOAA National Centers for Environmental Information}.
\newblock
  \url{https://data.noaa.gov/metaview/page?xml=NOAA/NESDIS/NGDC/MGG/passive_acoustic//iso/xml/DCLDE_2013.xml&view=getDataView}.

\bibitem[Gelfand et~al., 1996]{Gelfand1996}
Gelfand, A.~E., Sahu, S.~K., and Carlin, B.~P. (1996).
\newblock Efficient parametrizations for generalized linear mixed models.
\newblock {\em Bayesian Statistics 5: Proceedings of the Fifth Valencia
  International Meeting J M Bernardo (ed.) et al.}, pages 165--180.

\bibitem[Gillespie, 2004]{gillespie2004detection}
Gillespie, D. (2004).
\newblock Detection and classification of right whale calls using an `edge'
  detector operating on a smoothed spectrogram.
\newblock {\em Canadian Acoustics}, 32(2):39--47.

\bibitem[Gillespie et~al., 2020]{gillespiePassiveAcousticMethods2020}
Gillespie, D., Palmer, L., Macaulay, J., Sparling, C., and Hastie, G. (2020).
\newblock Passive acoustic methods for tracking the {{3D}} movements of small
  cetaceans around marine structures.
\newblock {\em PLOS ONE}, 15(5):e0229058.

\bibitem[Hatch et~al., 2016]{hatchCanYouHear2016}
Hatch, L.~T., Wahle, C.~M., Gedamke, J., Harrison, J., Laws, B., Moore, S.~E.,
  Stadler, J.~H., and Van~Parijs, S.~M. (2016).
\newblock Can you hear me here? {{Managing}} acoustic habitat in {{US}} waters.
\newblock {\em Endanger. Species Res.}, 30:171--186.

\bibitem[Helble et~al., 2016]{helble2016swim}
Helble, T.~A., Henderson, E.~E., Ierley, G.~R., and Martin, S.~W. (2016).
\newblock Swim track kinematics and calling behavior attributed to bryde's
  whales on the navy's pacific missile range facility.
\newblock {\em The Journal of the Acoustical Society of America},
  140(6):4170--4177.

\bibitem[Johnson and Tyack, 2003]{johnsonDigitalAcousticRecording2003}
Johnson, M.~P. and Tyack, P.~L. (2003).
\newblock A digital acoustic recording tag for measuring the response of wild
  marine mammals to sound.
\newblock {\em IEEE J. Oceanic Eng.}, 28(1):3--12.

\bibitem[Knowlton et~al., 2022]{knowltonFishingGearEntanglement2022}
Knowlton, A.~R., Clark, J.~S., Hamilton, P.~K., Kraus, S.~D., Pettis, H.~M.,
  Rolland, R.~M., and Schick, R.~S. (2022).
\newblock Fishing gear entanglement threatens recovery of critically endangered
  {{North Atlantic}} right whales.
\newblock {\em Conservation Science and Practice}, 4(8):e12736.

\bibitem[Kraus and Rolland, 2007]{krausUrbanWhaleNorth2007}
Kraus, S.~D. and Rolland, R.~M. (2007).
\newblock {\em The {{Urban Whale}}: {{North Atlantic Right Whales}} at the
  {{Crossroads}}}.
\newblock {Harvard University Press}.

\bibitem[Lewis and Shedler, 1979]{Lewis1979}
Lewis, P.~A. and Shedler, G.~S. (1979).
\newblock {Simulation of nonhomogeneous Poisson processes by thinning}.
\newblock {\em Naval research logistics quarterly}, 26:403--413.

\bibitem[Linden, 2023]{linden2023}
Linden, D. (2023).
\newblock Population size estimation of {North Atlantic} right whales from
  1990-2022. {US Dept. Commer. Northeast Fish. Sci. Cent. Tech Memo 314}.
\newblock Technical report, NOAA Fisheries.

\bibitem[Linderman and Adams, 2014]{lindermanDiscoveringLatentNetwork2014}
Linderman, S. and Adams, R. (2014).
\newblock Discovering {{Latent Network Structure}} in {{Point Process Data}}.
\newblock In {\em Proceedings of the 31st {{International Conference}} on
  {{Machine Learning}}}, pages 1413--1421. {PMLR}.

\bibitem[Matthews et~al., 2001]{matthews2001vocalisation}
Matthews, J., Brown, S., Gillespie, D., Johnson, M., McLanaghan, R., Moscrop,
  A., Nowacek, D., Leaper, R., Lewis, T., Tyack, P., et~al. (2001).
\newblock Vocalisation rates of the north atlantic right whale (eubalaena
  glacialis).
\newblock {\em J. Cetacean Res. Manage.}, 3(3):271--282.

\bibitem[Matthews and Parks, 2021]{matthewsOverviewNorthAtlantic2021}
Matthews, L.~P. and Parks, S.~E. (2021).
\newblock An overview of {{North Atlantic}} right whale acoustic behavior,
  hearing capabilities, and responses to sound.
\newblock {\em Marine Pollution Bulletin}, 173:113043.

\bibitem[Mayo et~al., 2018]{mayoDistributionDemographyBehavior2018}
Mayo, C.~A., Ganley, L., Hudak, C.~A., Brault, S., Marx, M.~K., Burke, E., and
  Brown, M.~W. (2018).
\newblock Distribution, demography, and behavior of {{North Atlantic}} right
  whales ( {{Eubalaena}} glacialis ) in {{Cape Cod Bay}}, {{Massachusetts}},
  1998-2013 : {{Right Whales}} in {{Cape Cod Bay}}.
\newblock {\em Mar. Mamm. Sci.}, 34(4):979--996.

\bibitem[Mellinger et~al., 2007]{mellingerOverviewFixedPassive2007}
Mellinger, D.~K., Stafford, K.~M., Moore, S.~E., Dziak, R.~P., and Matsumoto,
  H. (2007).
\newblock An {{Overview}} of {{Fixed Passive Acoustic Observation Methods}} for
  {{Cetaceans}}.
\newblock {\em Oceanography}, 20(4):36--45.

\bibitem[{Meyer-Gutbrod} et~al.,
  2023]{meyer-gutbrodRedefiningNorthAtlantic2023}
{Meyer-Gutbrod}, E.~L., Davies, K. T.~A., Johnson, C.~L., Plourde, S.,
  Sorochan, K.~A., Kenney, R.~D., Ramp, C., Gosselin, J.-F., Lawson, J.~W., and
  Greene, C.~H. (2023).
\newblock Redefining {{North Atlantic}} right whale habitat-use patterns under
  climate change.
\newblock {\em Limnology and Oceanography}, 68(S1):S71--S86.

\bibitem[Moore et~al., 2006]{mooreListeningLargeWhales2006}
Moore, S.~E., Stafford, K.~M., Mellinger, D.~K., and Hildebrand, J.~A. (2006).
\newblock Listening for {{Large Whales}} in the {{Offshore Waters}} of
  {{Alaska}}.
\newblock {\em Bioscience}, 56(1):49--55.

\bibitem[Nicvert et~al., 2024]{Nicvert2024}
Nicvert, L., Donnet, S., Keith, M., Peel, M., Somers, M.~J., Swanepoel, L.~H.,
  Venter, J., Fritz, H., and Dray, S. (2024).
\newblock {Using the multivariate Hawkes process to study interactions between
  multiple species from camera trap data}.
\newblock {\em Ecology}.

\bibitem[Palmer et~al., 2022]{palmerAccountingLombardEffect2022}
Palmer, K.~J., Wu, G.-M., Clark, C., and Klinck, H. (2022).
\newblock Accounting for the {{Lombard}} effect in estimating the probability
  of detection in passive acoustic surveys: {{Applications}} for single sensor
  mitigation and monitoring.
\newblock {\em The Journal of the Acoustical Society of America},
  151(1):67--79.

\bibitem[Parks et~al., 2019]{parksAcousticCrypsisCommunication2019}
Parks, S.~E., Cusano, D.~A., Van~Parijs, S.~M., and Nowacek, D.~P. (2019).
\newblock Acoustic crypsis in communication by {{North Atlantic}} right whale
  mother-calf pairs on the calving grounds.
\newblock {\em Biol. Lett.}, 15(10):20190485.

\bibitem[Parks et~al., 2010]{parksIndividualRightWhales2010}
Parks, S.~E., Johnson, M., Nowacek, D., and Tyack, P.~L. (2010).
\newblock Individual right whales call louder in increased environmental noise.
\newblock {\em Biology Letters}, 7(1):33--35.

\bibitem[Parks et~al., 2011]{parksSoundProductionBehavior2011}
Parks, S.~E., Searby, A., C{\'e}l{\'e}rier, A., Johnson, M.~P., Nowacek, D.~P.,
  and Tyack, P.~L. (2011).
\newblock Sound production behavior of individual {{North Atlantic}} right
  whales: Implications for passive acoustic monitoring.
\newblock {\em Endangered Species Research}, 15(1):63--76.

\bibitem[Parks and Tyack, 2005]{parks2005sound}
Parks, S.~E. and Tyack, P.~L. (2005).
\newblock Sound production by {{North Atlantic}} right whales ({{Eubalaena}}
  glacialis) in surface active groups.
\newblock {\em The Journal of the Acoustical Society of America},
  117(5):3297--3306.

\bibitem[Parks et~al., 2009]{parksVariabilityAmbientNoise2009}
Parks, S.~E., Urazghildiiev, I., and Clark, C.~W. (2009).
\newblock Variability in ambient noise levels and call parameters of {{North
  Atlantic}} right whales in three habitat areas.
\newblock {\em J. Acoust. Soc. Am.}, 125(2):1230--1239.

\bibitem[Payne and McVay, 1971]{payne1971songs}
Payne, R.~S. and McVay, S. (1971).
\newblock Songs of humpback whales: Humpbacks emit sounds in long, predictable
  patterns ranging over frequencies audible to humans.
\newblock {\em Science}, 173(3997):585--597.

\bibitem[Radford et~al., 2014]{radfordAcousticCommunicationNoisy2014}
Radford, A.~N., Kerridge, E., and Simpson, S.~D. (2014).
\newblock Acoustic communication in a noisy world: Can fish compete with
  anthropogenic noise?
\newblock {\em Behavioral Ecology}, 25(5):1022--1030.

\bibitem[Rathbun, 1996]{Rathbun1996}
Rathbun, S.~L. (1996).
\newblock {Asymptotic properties of the maximum likelihood estimator for
  spatio-temporal point processes}.
\newblock {\em Journal of Statistical Planning and Inference}, 51:55--74.

\bibitem[Reinhart, 2018]{Reinhart2018}
Reinhart, A. (2018).
\newblock {A review of self-exciting spatio-temporal point processes and their
  applications}.
\newblock {\em Statistical Science}, 33:299--318.

\bibitem[Richardson et~al., 1995]{richardson1995marine}
Richardson, W.~J., Greene~Jr, C.~R., Malme, C.~I., and Thomson, D.~H. (1995).
\newblock {\em Marine Mammals and Noise}.
\newblock Academic Press.

\bibitem[{Root-Gutteridge} et~al.,
  2018]{root-gutteridgeLifetimeChangingCalls2018}
{Root-Gutteridge}, H., Cusano, D.~A., Shiu, Y., Nowacek, D.~P., Van~Parijs,
  S.~M., and Parks, S.~E. (2018).
\newblock A lifetime of changing calls: {{North Atlantic}} right whales,
  {{Eubalaena}} glacialis, refine call production as they age.
\newblock {\em Anim. Behav.}, 137:21--34.

\bibitem[Schevill and Watkins, 1962]{schevill1962photograph}
Schevill, W.~E. and Watkins, W.~A. (1962).
\newblock A photograph record.
\newblock Technical report, Woods Hole Oceanographic Institution. Woods Hole,
  Mass.

\bibitem[Schliep et~al., 2023]{schliep2023assessing}
Schliep, E.~M., Gelfand, A.~E., Clark, C.~W., Mayo, C.~M., McKenna, B., Parks,
  S.~E., Yack, T.~M., and Schick, R.~S. (2023).
\newblock {Assessing marine mammal abundance: a novel data fusion}.
\newblock {\em arXiv preprint arXiv:2310.08397}.

\bibitem[Tyack and Clark, 2000]{tyack2000communication}
Tyack, P.~L. and Clark, C.~W. (2000).
\newblock Communication and acoustic behavior of dolphins and whales.
\newblock In {\em Hearing by whales and dolphins}, pages 156--224. Springer.

\bibitem[Urazghildiiev, 2014]{urazghildiievStatisticalAnalysisNorth2014}
Urazghildiiev, I.~R. (2014).
\newblock Statistical analysis of {{North Atlantic}} right whale ({{Eubalaena}}
  glacialis) signal trains in {{Cape Cod Bay}}, spring 2012.
\newblock {\em The Journal of the Acoustical Society of America},
  136(5):2851--2860.

\bibitem[Vanderlaan and Taggart, 2007]{vanderlaan2007}
Vanderlaan, A. S.~M. and Taggart, C.~T. (2007).
\newblock Vessel colisions with whales: The probability of lethal injury based
  on vessel speed.
\newblock {\em Mar. Mamm. Sci.}, 23(1):144--156.

\bibitem[Watkins and Schevill, 1972]{watkinsSoundSourceLocation1972a}
Watkins, W.~A. and Schevill, W.~E. (1972).
\newblock Sound source location by arrival-times on a non-rigid
  three-dimensional hydrophone array.
\newblock {\em Deep Sea Research and Oceanographic Abstracts}, 19(10):691--706.

\bibitem[White and Gelfand, 2021]{White2021}
White, P.~A. and Gelfand, A.~E. (2021).
\newblock {Generalized evolutionary point processes: model specifications and
  model comparison}.
\newblock {\em Methodology and Computing in Applied Probability},
  23:1001--1021.

\end{thebibliography}

\end{document}


\def\spacingset#1{\renewcommand{\baselinestretch}%
{#1}\small\normalsize} \spacingset{1}


\if1\blind
{
  \title{\bf Supplementary Material for ``Analyzing whale calling through Hawkes process modeling''}
  \author{Bokgyeong Kang, Erin M. Schliep, Alan E. Gelfand, Tina M. Yack,\\
  Christopher W. Clark, and Robert S. Schick\\
  }
  \maketitle
} \fi

\if0\blind
{
  \bigskip
  \bigskip
  \bigskip
  \begin{center}
    {\LARGE\bf Supplementary Material for ``Analyzing whale calling through Hawkes process modeling''}
\end{center}
  \medskip
} \fi

\spacingset{1.9} 





\section{Multi-channel CCB data}
\label{sup:sec:mdata}

Table~\ref{sup:tab:distMARU} shows the pairwise distances between MARUs (km).

\setlength{\tabcolsep}{5pt}
\begin{table}[!ht]
\caption{Distances between MARUs (km). \label{sup:tab:distMARU}}
\begin{center}
\begin{tabular}{lrrrrrrrrr}
  \toprule
    & 2 & 3 & 4 & 5 & 6 & 7 & 8 & 9 & 10 \\ 
  \midrule
  1 &  7.6 & 13.7 & 16.3 & 16.1 & 10.3 &  8.7 &  8.3 & 18.0 & 15.2 \\ 
  2 &  &  7.1 & 13.1 & 16.3 & 14.7 &  8.0 & 15.8 & 21.6 & 21.4 \\ 
  3 &  &  &  8.3 & 14.3 & 16.8 &  8.4 & 20.8 & 21.8 & 24.4 \\ 
  4 &  &  &  &  7.2 & 13.8 &  7.6 & 20.7 & 15.9 & 21.2 \\ 
  5 &  &  &  &  &  9.1 &  8.5 & 17.4 &  8.8 & 15.4 \\ 
  6 &  &  &  &  &  &  8.4 &  8.5 &  7.8 &  7.7 \\ 
  7 &  &  &  &  &  &  & 13.5 & 14.0 & 16.0 \\ 
  8 &  &  &  &  &  &  &  & 15.0 &  8.5 \\ 
  9 &  &  &  &  &  &  &  &  &  8.9 \\ 
   \bottomrule
\end{tabular}    
\end{center}
\end{table}




\section{Inference}
\label{sup:sec:infer}

\subsection{The likelihood}
\label{sup:subsec:likelihood}

The likelihood function for Model (iv), NHPP+GP+CC, is given by
\begin{align*}
    L(\btheta \mid \mathcal{T}) = \exp \left\{ - \sum_{k=1}^K \int_{0}^T \lambda_k(t \mid \mathcal{H}_{t}; \btheta) dt \right\} \prod_{i=1}^n \lambda_{m_i}(t_i \mid \mathcal{H}_{t_i}; \btheta).
\end{align*}
Estimation for this model can be viewed as an incomplete data problem. \citet{Veen2008} introduced a latent variable $z_i$ approach which, applied to our setting, indicates whether the call at $t_i$ is a contact call or is a countercall caused by a previous call, i.e.,
\begin{align}
    z_i = \left\{ \begin{array}{ll}
        0 & \mbox{if the call at $t_i$ is a contact call}\\
        j & \mbox{if the call at $t_i$ is a countercall caused by a previous call at $t_j$}
    \end{array}\right. \label{sup:eq:zi}
\end{align}
for $i = 1,\dots,n$ and $j \in \{r: t_r < t_i\}$. Given that the branching structure $\bz= \{z_1, \dots, z_n\}$ is known, we can partition the observed calls into $n$ sets $\mathcal{T}_0, \dots, \mathcal{T}_{n-1}$ where $\mathcal{T}_j = \{t_i; z_i = j\}$ for $0 \leq j \leq n$. That is, $\mathcal{T}_0$ is the set of contact calls and $\mathcal{T}_j$ is the set of countercalls excited by call $j$ at $t_j$. These $n$ sets are mutually exclusive and their union contains all the observed calls. We can see that $\mathcal{T}_0$ is the collection of calls generated by a NHPP with intensity $\mu_k(t)$ across $k$, and $\mathcal{T}_j$ for $j > 0$ is the collection of calls generated by a NHPP with intensity $\alpha_{m_j} e^{-\eta(t-t_j)} e^{-\phi d_{m_j, k}}$ across $k$. Multiplying together the likelihoods from each of these processes, we obtain the complete-data likelihood given by
\begin{align}
    &L(\btheta \mid \mathcal{T}, \bz) = \exp \left\{ - \sum_{k=1}^K \int_0^T \mu_k(t) dt \right\}
    \prod_{i=1}^n \mu_{m_i}(t_i)^{\mathbb{I}(z_i = 0)} \label{sup:eq:compLik} \\
    &\times \exp \left\{ -\sum_{k=1}^K \sum_{j: t_j < t} \alpha_k e^{- \phi d_{m_j,k}} \int_{t_j}^T  e^{-\eta(t-t_j)}  dt \right\} \prod_{i=1}^n \prod_{j: t_j < t_i} \left[ \alpha_{m_j} e^{-\eta(t_i-t_j)} e^{- \phi d_{m_j,k}} \right] ^{\mathbb{I}(z_i = j)}. \nonumber
\end{align}

\subsection{Likelihood approximation}
\label{sup:subsec:likAppx}

The evaluation of the likelihood \eqref{sup:eq:compLik} is intractable because $\int_0^T \mu_k(t) dt$ has no closed-form expression. We approximate the integral using linear interpolation. We discretize the observed time window $(0, T]$ into $M+1$ subintervals with partition points $\mathcal{P} = \{0, \frac{T}{M}, \frac{2T}{M}, \dots, T\}$. Let $\Tilde{\bmu}_k = \left( \mu_k(0), \mu_k(\frac{T}{M}), \dots, \mu_k(T) \right)^\top$ be the background intensity evaluated at $\mathcal{P}$. We can create the associated design matrix. We evaluate the GP at the partition points $\mathcal{P}$ which is given by $\Tilde{\bw} = \left( w(0), w(\frac{T}{M}), \dots, w(T) \right)^\top$. 
We define $\Tilde{\mu}_k(t)$ as the linear interpolation on $(\mathcal{P}, \Tilde{\bmu}_k)$ and replace $\mu_k(t)$ in \eqref{sup:eq:compLik} with $\Tilde{\mu}_k(t)$. The linear interpolation $\Tilde{\mu}_k(t)$ provides a good approximation to $\mu_k(t)$ for large enough $M$, and $\int_0^T \Tilde{\mu}_k(t) dt$ is easy to compute using the trapezoidal rule. We discretize the observed time window every 20 minutes, i.e., $\frac{T}{M} = 20$, for both simulation an real data.

\subsection{Bayesian model fitting}
\label{sup:subsec:fitting}

We fit our models within the Bayesian framework. For $\Tilde{\beta}_j$ and $\log(\Tilde{\delta})$, we assume a conjugate normal prior with mean 0 and variance 100. We assign a conjugate inverse-gamma prior with shape 2 and scale 1 for the variance parameters $\tau_j$ for $j = 0,1, \dots, p$ and $\tau_{\delta}$. For the temporal decay parameter $\eta$, we assume a Uniform$\big(\frac{3}{20}$, $\frac{3}{\min_{i:t_i \in \mathcal{T}}\{\vert t_{i}-t_{i+1} \vert\}}\big)$ prior. This reflects our belief that, at most, a whale will take 20 minutes to respond to a contact call. For the spatial decay parameter $\phi$, we assume a Uniform$\big(\frac{3}{ \max_{k, \ell}\{d_{k,\ell}\} }$, $\frac{3}{ \min_{k, \ell}\{d_{k,\ell}\}} \big) $ prior to entirely cover the array of MARUs. We conducted a simulation experiment for choosing a suitable prior distribution for the countercall intensity parameter $\alpha$. Let Gamma$(a,b)$ denote a gamma distribution with shape $a$ and scale $b$. We compared Gamma(2,1), Gamma(0.001, 1000), and spike and slab priors. We found that using the Gamma(0.001, 1000) prior recovers truth well for various values of $\alpha \geq 0$.

We use Markov chain Monte Carlo (MCMC) for model fitting and generate posterior samples of the latent variable $\bz$ and model parameters $\btheta$ as follows:

\begin{enumerate}[1.]
    \item \textbf{Sampling a latent variable $z_i$.} The full conditional posterior probability of $z_i = j$ is given by 
    \begin{align*}
        P(z_i = j \mid \text{rest}) = \left\{
        \begin{array}{ll}
            \frac{\Tilde{\mu}_{m_i}(t_i)}{\Tilde{\mu}(t_i) + \sum_{r=1}^{i-1} \alpha_{m_r} e^{-\eta(t_i - t_r)} e^{-\phi d_{m_r, m_i}} } & \text{if } j = 0 \\
            \frac{ \alpha_{m_j} e^{-\eta(t_i - t_j)} e^{-\phi d_{m_j, m_i}} }{\Tilde{\mu}(t_i) + \sum_{r=1}^{i-1} \alpha_{m_r} e^{-\eta(t_i - t_r)}e^{-\phi d_{m_r, m_i}} } & \text{if } j = 1,\cdots,i-1 .
        \end{array}
        \right.
    \end{align*}
    That is, $z_i$ has a multinomial distribution with probability $P(z_i = j \mid \text{rest})$ for $i = 2,\dots,n$ and $j = 1,\cdots,i-1$.
    \item \textbf{Sampling covariate coefficients $\bbeta_j$.} For $j = 0, 1, \dots, p$ the full conditional posterior of $\bbeta_j$ is given by
    \begin{align*}
        \pi(\bbeta_j \mid \text{rest}) \propto \exp\left\{  - \sum_{k=1}^K  \int_{0}^T \Tilde{\mu}_k(t) dt \right\}  \prod_{i=1}^n \Tilde{\mu}_{m_i}(t_i)^{\mathbb{I}(z_i = 0)} p(\bbeta_j),
    \end{align*}
    where $p(\bbeta_j)$ is the prior. Although there is no conjugate prior, it is straightforward to use a Metropolis--Hastings (MH) update.
    \item \textbf{Sampling $\Tilde{\beta}_j$.} For $j = 0, 1, \dots, p$ the full conditional posterior of $\Tilde{\beta}_j$ follows a normal distribution with mean $m_{\Tilde{\beta}_j}$ and variance $s_{\Tilde{\beta}_j}$ where
    \begin{align*}
        s_{\Tilde{\beta}_j} &= \frac{1}{(1^{\top} V^{-1} 1)/\tau_j + 1/100}\\
        m_{\Tilde{\beta}_j} &= s_{\Tilde{\beta}_j} (\bbeta_j^{\top} V^{-1} 1)/\tau_j.
    \end{align*}
    \item \textbf{Sampling $\tau_j$.} For $j = 0, 1, \dots, p$, the full conditional posterior for  $\tau_j$ is given by
    \begin{align*}
        \pi(\tau_j \mid \text{rest}) & \propto \det(\tau_j V)^{-1/2} \exp\left\{ - \frac{1}{2} (\bbeta_j - \Tilde{\beta}_j)^\top (\tau_j V)^{-1} (\bbeta_j - \Tilde{\beta}_j) \right\} p(\tau_j),
    \end{align*}
    where $p(\tau_j)$ is a conjugate inverse-gamma prior with shape $a_\tau$ and scale $b_\tau$. The full conditional posterior for $\tau_j$ follows a inverse-gamma with shape $= a_\tau + K/2$ and scale $= b_\tau + \frac{1}{2} (\bbeta_j - \Tilde{\beta}_j)^\top V^{-1} (\bbeta_j - \Tilde{\beta}_j)$.
    \item \textbf{Sampling Gaussian process $\Tilde{\bw}$.} The full conditional posterior for $\Tilde{\bw}$ is given by
    \begin{align*}
        \pi(\Tilde{\bw} \mid \text{rest}) \propto \exp\left\{  - \sum_{k=1}^K  \int_{0}^T \Tilde{\mu}_k(t) dt \right\}  \prod_{i=1}^n \Tilde{\mu}_{m_i}(t_i)^{\mathbb{I}(z_i = 0)} p(\Tilde{\bw}),
    \end{align*}
    where $p(\Tilde{\bw})$ is the prior of $\Tilde{\bw}$. MCMC is known to perform poorly for Gaussian process models due to the strong correlation among variables. We use elliptical slice sampling \citep{Murray2010} for updating $\Tilde{\bw}$. This algorithm is simple and more efficient than other methods.
    \item \textbf{Sampling GP coefficients $\log(\bdelta)$.} The full conditional posterior for $\log(\bdelta)$ is given by
    \begin{align*}
        \pi(\log(\bdelta) \mid \text{rest}) \propto \exp\left\{  - \sum_{k=1}^K  \int_{0}^T \Tilde{\mu}_k(t) dt \right\}  \prod_{i=1}^n \Tilde{\mu}_{m_i}(t_i)^{\mathbb{I}(z_i = 0)} p(\log(\bdelta)),
    \end{align*}
    where $p(\log(\bdelta))$ is the prior. We use a MH update.
    \item \textbf{Sampling $\Tilde{\delta}$.} The full conditional posterior of $\Tilde{\delta}$ follows a normal distribution with mean $m_{\Tilde{\delta}}$ and variance $s_{\Tilde{\delta}}$ where
    \begin{align*}
        s_{\Tilde{\delta}} &= \frac{1}{(1^{\top} V^{-1} 1)/\tau_{\delta} + 1/100}\\
        m_{\Tilde{\delta}} &= s_{\Tilde{\delta}} (\bdelta^{\top} V^{-1} 1)/\tau_{\delta}.
    \end{align*}
    \item \textbf{Sampling $\tau_\delta$.} The full conditional posterior for  $\tau_\delta$ is given by
    \begin{align*}
        \pi(\tau_\delta \mid \text{rest}) & \propto \det(\tau_\delta V)^{-1/2} \exp\left\{ - \frac{1}{2} (\bdelta - \Tilde{\delta})^\top (\tau_\delta V)^{-1} (\bdelta - \Tilde{\delta}) \right\} p(\tau_\delta),
    \end{align*}
    where $p(\tau_\delta)$ is a conjugate inverse-gamma prior with shape $a_\delta$ and scale $b_\delta$. The full conditional posterior for $\tau_j$ follows a inverse-gamma with shape $= a_\delta + K/2$ and scale $= b_\delta + \frac{1}{2} (\bdelta - \Tilde{\delta})^\top V^{-1} (\bdelta - \Tilde{\delta})$.
    \item \textbf{Sampling excitement intensity $\alpha_{\ell}$.} The full conditional posterior for $\alpha_{\ell}$ is given by
    \begin{align*}
        p(\alpha_{\ell} \mid \text{rest}) \propto 
        &\exp \left\{ - \alpha_{\ell} \sum_{j=1}^n \sum_{k=1}^K \mathbb{I}(m_j = \ell) e^{-\phi d_{m_j, k}} \int_{t_i}^T e^{-\eta (t - t_j)} dt \right\}\\
        &\times  \alpha_{\ell}^{\sum_{i=1}^n \sum_{j=1}^{i-1} \mathbb{I}(z_i = j, m_j = \ell)} p(\alpha),
    \end{align*}
    $p(\alpha)$ is the conjugate gamma prior with shape $a_{\alpha}$ and scale $b_{\alpha}$. The full conditional posterior of $\alpha_{\ell}$ follows a gamma with shape $= a_{\alpha} + \sum_{i=1}^n \sum_{j=1}^{i-1} \mathbb{I}(z_i = j, m_j = \ell)$ and scale $= \big( 1/b_{\alpha} + \sum_{j=1}^n \sum_{k=1}^K \mathbb{I}(m_j = \ell) \int_{t_i}^T e^{-\eta (t - t_j)} e^{-\phi d_{m_j, k}} dt \big)^{-1}$.
    \item \textbf{Sampling temporal decay $\eta$.} The posterior distribution for $\eta$ is given by
    \begin{align*}
        p(\eta \mid \alpha, \bz, \mathcal{T}) \propto &\exp \left\{ - \sum_{j=1}^n \sum_{k=1}^K \alpha_{m_j} e^{-\phi d_{m_j, k}} \int_{t_i}^T e^{-\eta (t - t_j)} dt \right\}\\
        & e^{-\eta \sum_{i=1}^n \sum_{j=1}^{i-1} (t_i - t_j) \mathbb{I}(z_i = j) } p(\eta),
    \end{align*}
    where $p(\eta)$ is a prior. There is no conjugate prior so we use a MH update.
    \item \textbf{Sampling spatial decay $\phi$.} The posterior distribution for $\eta$ is given by
    \begin{align*}
        p(\eta \mid \alpha, \bz, \mathcal{T}) \propto &\exp \left\{ - \sum_{j=1}^n \sum_{k=1}^K \alpha_{m_j} e^{-\phi d_{m_j, k}} \int_{t_i}^T e^{-\eta (t - t_j)} dt \right\}\\
        & e^{-\phi \sum_{i=1}^n \sum_{j=1}^{i-1} d_{m_j, m_i} \mathbb{I}(z_i = j) } p(\phi),
    \end{align*}
    where $p(\phi)$ is a prior. There is no conjugate prior so we use a MH update.
\end{enumerate}

\section{Simulation study}
\label{sup:sec:sim}

\subsection{Single-channel data}
\label{sup:subsec:usim}

We simulate a dataset from each of the four models (i) NHPP, (ii) NHPP+GP, (iii) NHPP+CC, and (iv) NHPP+GP+CC. We fix $K = 1$ and consider 5 consecutive days for the observed time window. For all but $\beta_0$, the true values of the model parameters are set to the estimated posterior means from Model (iv) fitted to the single-channel dataset (as discussed in Section 6.1). We choose $\beta_0$ such that the total number of calls is approximately equal to what is observed in the real data. 

\begin{figure}[!b]
\begin{center}
\includegraphics[width=0.8\textwidth]{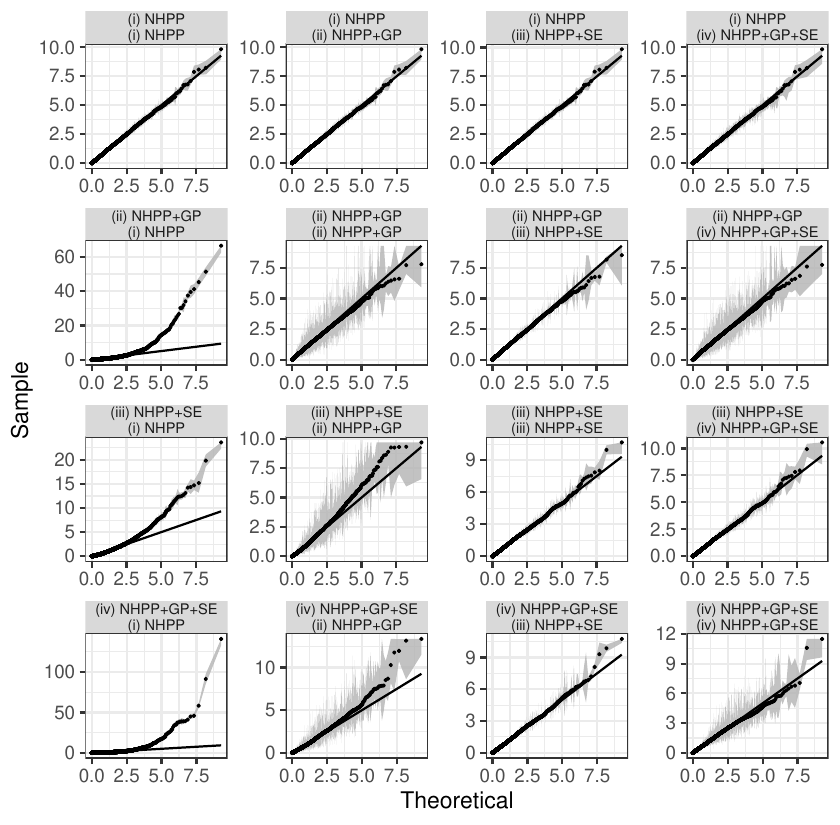}
\end{center}
\caption{Q-Q plots for $\hat{d}^{\ast}_i$'s against an Exp(1) distribution. Gray shades represent 95\% credible bands. Top and bottom labels denote generating and fitting models, respectively. \label{sup:fig:usimQQband}}
\end{figure}

We fit Models (i) to (iv) to each of the four simulated datasets using MCMC run for 100,000 iterations. The first 10,000 posterior samples are discarded as burn-in and the remaining 90,000 are used for posterior inference. We examined trace plots of chains for convergence and no issues were detected. 

\setlength{\tabcolsep}{3pt}
\begin{table}[!t]
\caption{Mean squared differences between sample and theoretical quantiles stemming from Models (i) to (iv) fitted to single-channel data generated from each of the models. \label{sup:tab:usimMSD}}
\small
\begin{center}
\begin{tabular}{lrrrr}
  \toprule
  Generating $\backslash$ Fitting model & (i) NHPP & (ii) NHPP+GP & (iii) NHPP+CC & (iv) NHPP+GP+CC \\ 
  \midrule
  (i) NHPP & 0.000 & 0.000 & 0.000 & 0.000 \\ 
  (ii) NHPP+GP & 3.580 & 0.004 & 0.002 & 0.004 \\ 
  (iii) NHPP+SE & 0.331 & 0.028 & 0.002 & 0.002 \\ 
  (iv) NHPP+GP+SE & 9.703 & 0.038 & 0.002 & 0.006 \\ 
   \bottomrule
\end{tabular}
\end{center}
\end{table}

\setlength{\tabcolsep}{3pt}
\begin{table}[!b]
\caption{DIC from Models (i) to (iv) fitted to data generated from each of the models. All entries are of order $10^{3}$. Bolds denote the fitting models with the smallest DIC.
\label{sup:tab:usimDIC}}
\small
\begin{center}
\begin{tabular}{lrrrr}
  \toprule
Generating $\backslash$ Fitting & (i) NHPP & (ii) NHPP+GP & (iii) NHPP+CC & (iv) NHPP+GP+CC \\ 
  \midrule
(i) NHPP & \textbf{6.0} & \textbf{6.0} & \textbf{6.0} & \textbf{6.0} \\ 
  (ii) NHPP+GP & 8.0 & \textbf{1.7} & 2.4 & \textbf{1.7} \\ 
  (iii) NHPP+SE & 5.9 & 4.4 & \textbf{4.2} & \textbf{4.2} \\ 
  (iv) NHPP+GP+SE & 7.9 & 1.2 & 1.3 & \textbf{1.0} \\ 
   \bottomrule
\end{tabular}
\end{center}
\end{table}

We use RTCT and DIC to assess how well fitting models can distinguish generating models. Figure~\ref{sup:fig:usimQQband} displays the posterior mean empirical Q-Q plots for $d^{\ast}_i$'s against an Exp(1) distribution for each combination of generating and fitting models. Table~\ref{sup:tab:usimMSD} shows the mean squared difference (MSD) between sample and theoretical quantiles for each scenario. Model (i) only fits data generated under Model (i). Model (ii) fits data generated from the true model (ii) and its submodel (i). Models (iii) and (iv) seem to fit every dataset fairly well. Table~\ref{sup:tab:usimDIC} shows that DIC correctly selects the true model and model(s) that have the true model as a submodel.

\begin{figure}[!t]
\begin{center}
\includegraphics[width=\textwidth]{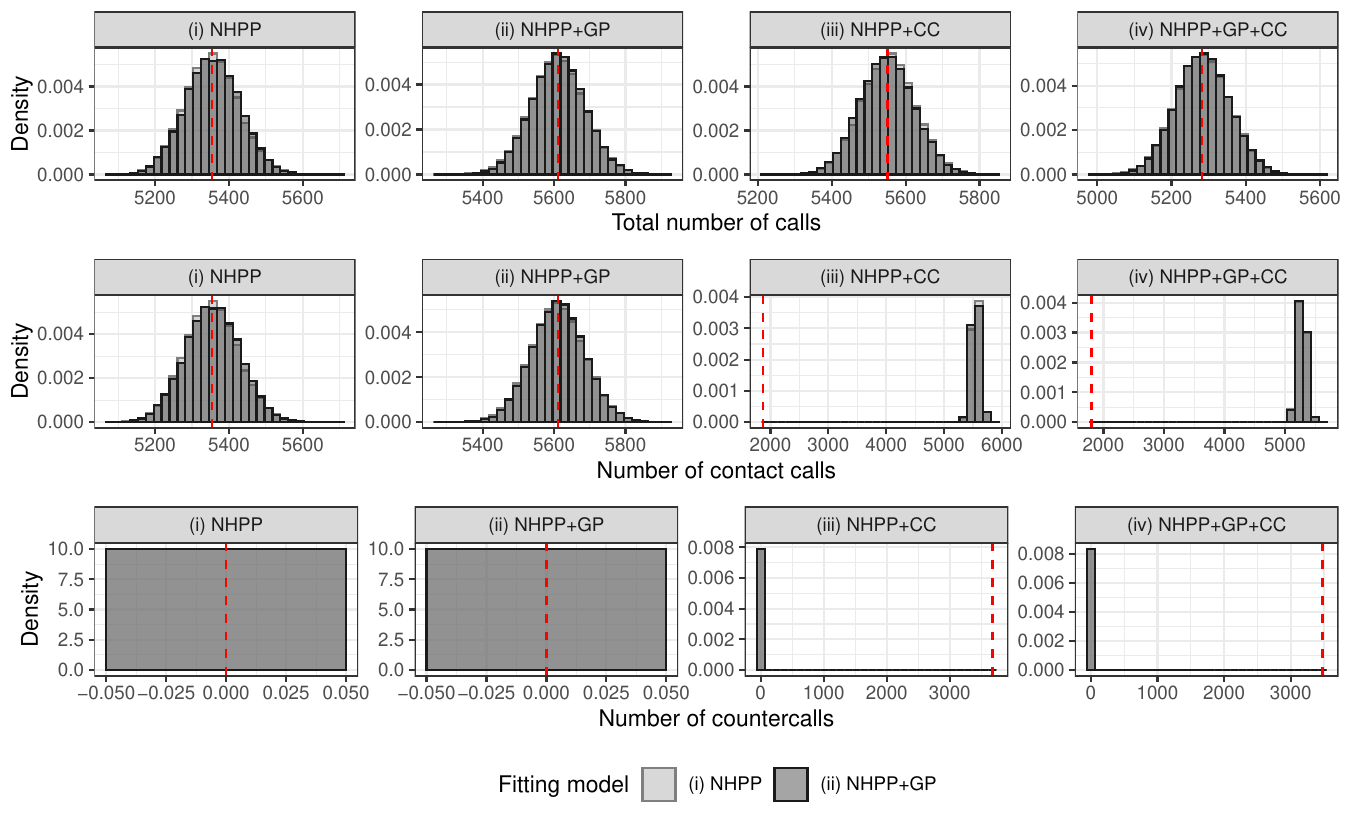}
\end{center}
\caption{Posterior distributions of the expected 
total number of calls (top row), the expected number of contact calls (middle row), and the expected number of countercalls (bottom call) from (i) NHPP (light gray) and (ii) NHPP+GP (dark gray). Facet labels denote generating models. Red dashed vertical lines are the observed values. \label{sup:fig:usimNum12}}
\end{figure}

\begin{figure}[!t]
\begin{center}
\includegraphics[width=\textwidth]{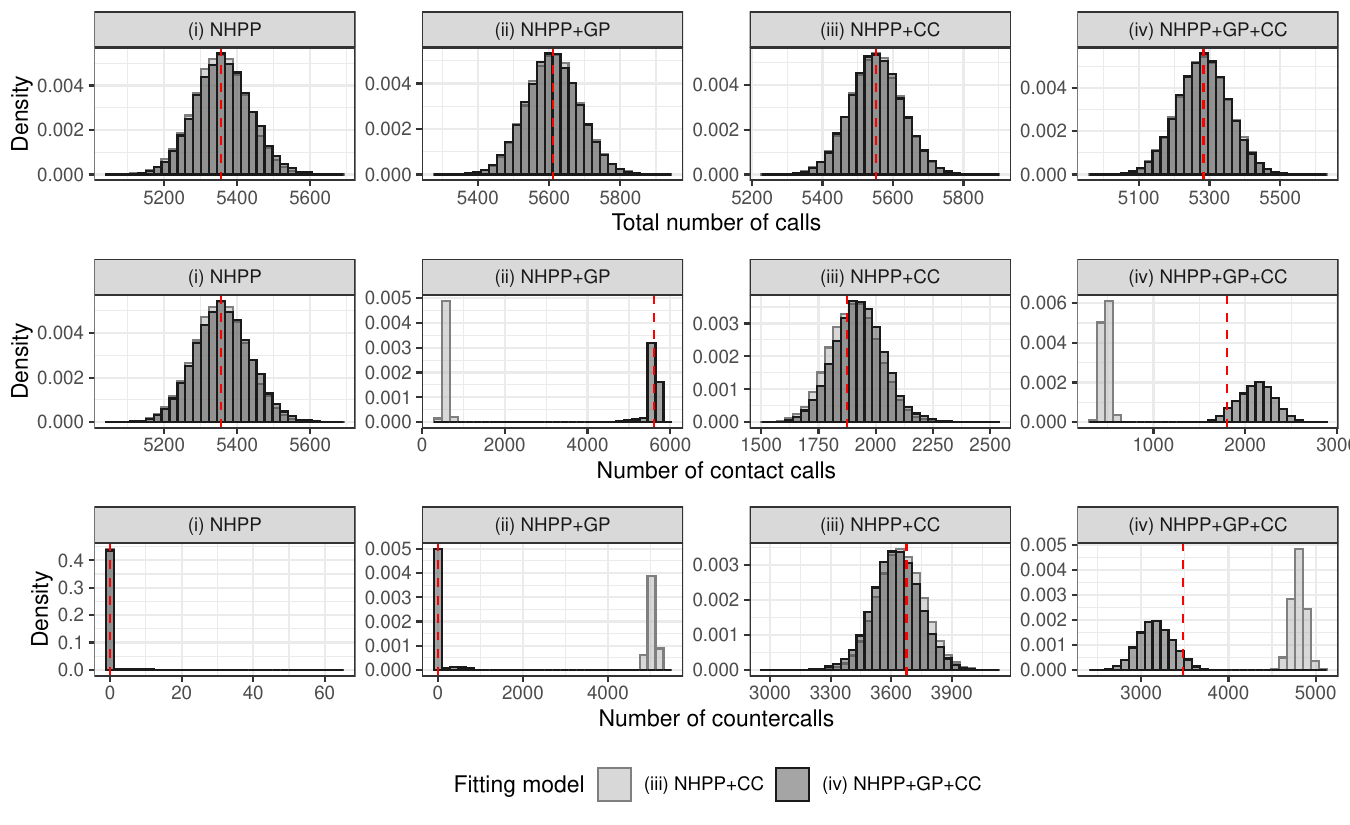}
\end{center}
\caption{Posterior distributions of the expected 
total number of calls (top row), the expected number of contact calls (middle row), and the expected number of countercalls (bottom call) from (iii) NHPP+CC (light gray) and (iv) NHPP+GP+CC (dark gray). Facet labels denote generating models. Red dashed vertical lines are the observed values. \label{sup:fig:usimNum34}}
\end{figure}

We estimate the expected total number of calls, expected number of contact calls, and expected number of countercalls for each case. Figure~\ref{sup:fig:usimNum12} displays the empirical posterior distributions for the expectations from Models (i) NHPP and (ii) NHPP+GP. Each model recovers the true expected total number of calls well for every dataset. When there are countercalls in the generated data, Models (i) and (ii), which have no CC component, overestimate the expected number of contact calls to recover the expected total number of calls. Figure~\ref{sup:fig:usimNum34} displays the empirical posterior distributions for the expectations from Models (iii) NHPP+CC and (iv) NHPP+GP+CC. Model (iii) recovers the simulated truth well for datasets generated from itself and its submodel, i.e., Model (i) and (iii). When there is a GP in the generative model, Model (iii) considerably underestimates the expected number of contact calls and overestimates the expected number of countercalls. This implies that the CC component of Model (iii) inappropriately accounts for variability from the GP component of the intensity. Model (iv) recovers the truth fairly well for every dataset. This implies that the GP component is well distinguished from the CC component of the intensity.

\subsection{Multi-channel data}
\label{sup:subsec:msim}

\setlength{\tabcolsep}{3pt}
\begin{table}[!t]
\caption{Mean squared differences between sample and theoretical quantiles stemming from Models (i) to (iv) fitted to multi-channel data generated from each of the models. \label{sup:tab:msimMSD}}
\small
\begin{center}
\begin{tabular}{lrrrr}
  \toprule
  Generating $\backslash$ Fitting model & (i) NHPP & (ii) NHPP+GP & (iii) NHPP+CC & (iv) NHPP+GP+CC \\ 
  \midrule
  (i) NHPP & 0.001 & 0.001 & 0.001 & 0.001 \\ 
  (ii) NHPP+GP & 0.182 & 0.008 & 0.011 & 0.008 \\ 
  (iii) NHPP+CC & 0.150 & 0.010 & 0.001 & 0.001 \\ 
  (iv) NHPP+GP+CC & 1.076 & 0.002 & 0.003 & 0.004 \\ 
   \bottomrule
\end{tabular}
\end{center}
\end{table}

\begin{figure}[!b]
\begin{center}
\includegraphics[width=0.9\textwidth]{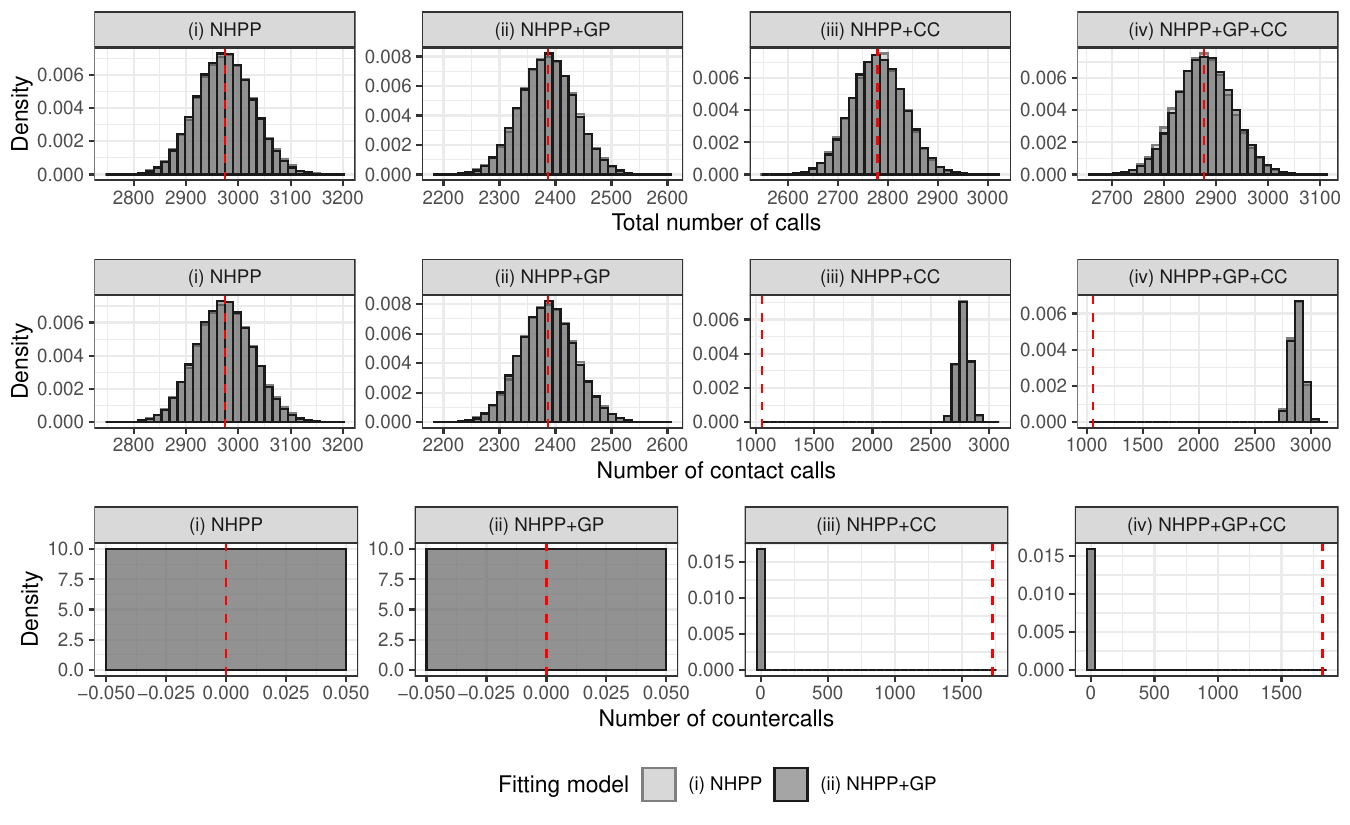}
\end{center}
\caption{Posterior distributions of the expected 
total number of calls (top row), the expected number of contact calls (middle row), and the expected number of countercalls (bottom call) from (i) NHPP (light gray) and (ii) NHPP+GP (dark gray). Facet labels denote generating models. Red dashed vertical lines are the observed values. \label{sup:fig:msimNum12}}
\end{figure}

We simulate a dataset from each of the four models (i) NHPP, (ii) NHPP+GP, (iii) NHPP+CC, and (iv) NHPP+GP+CC.
We use RTCT to assess model adequacy for each scenario. 
Table~\ref{sup:tab:msimMSD} displays the mean squared differences between sample and theoretical quantiles. This shows that Model (i) only fits data generated under Model (i). Models (ii) to (iv) seem to fit every dataset fairly well.

We estimate the expected total number of calls, expected number of contact calls, and expected number of countercalls. Figure~\ref{sup:fig:msimNum12} displays the empirical posterior distributions for the expectations from Models (i) NHPP and (ii) NHPP+GP. Each model recovers the true expected total number of calls well for every dataset. When there are countercalls in the generated data, Models (i) and (ii), which have no CC component, overestimate the expected number of contact calls to recover the expected total number of calls.

\section{Analysis of North Atlantic right whale call data}
\label{sup:sec:real}

\subsection{Multi-channel CCB data}
\label{sup:subsec:mreal}

\begin{figure}[!b]
\begin{center}
\includegraphics[width=\textwidth]{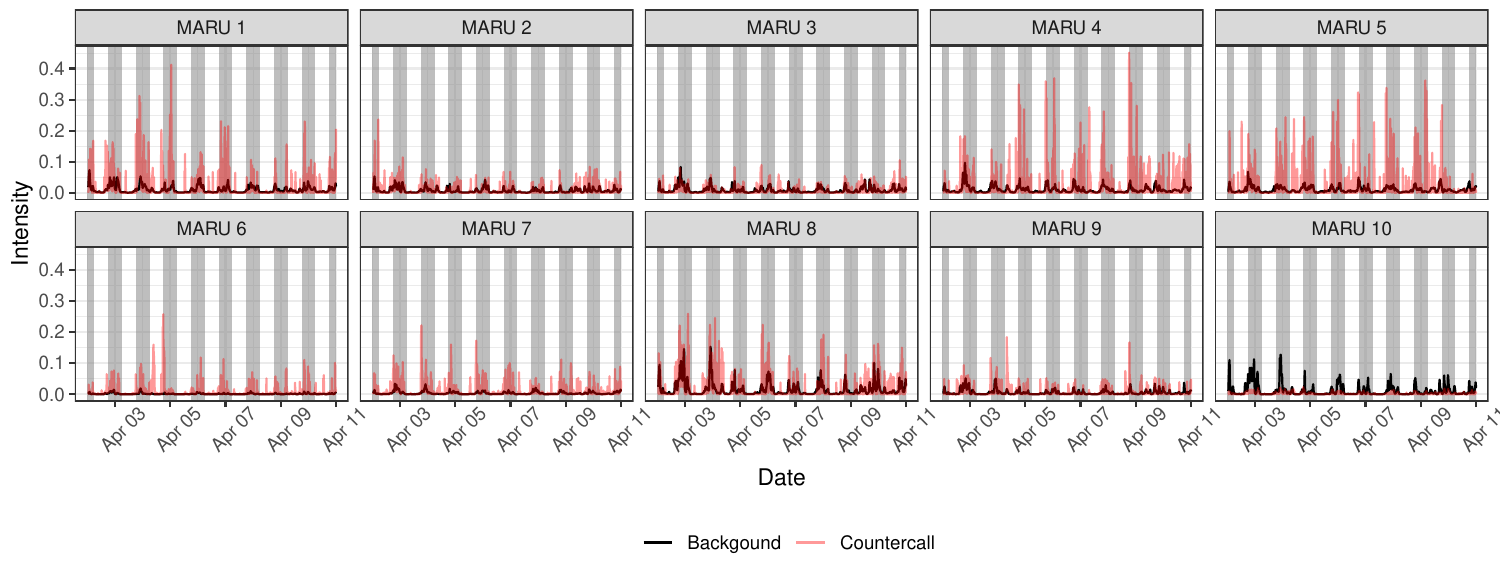}
\end{center}
\caption{Estimated background intensity (black) and countercall intensity (red) over time for each MARU. Gray shade indicates 6 p.m. to 6 a.m.\label{sup:fig:ccbLamBackSE}}
\end{figure}

\begin{figure}[!t]
\begin{center}
\includegraphics[width=\textwidth]{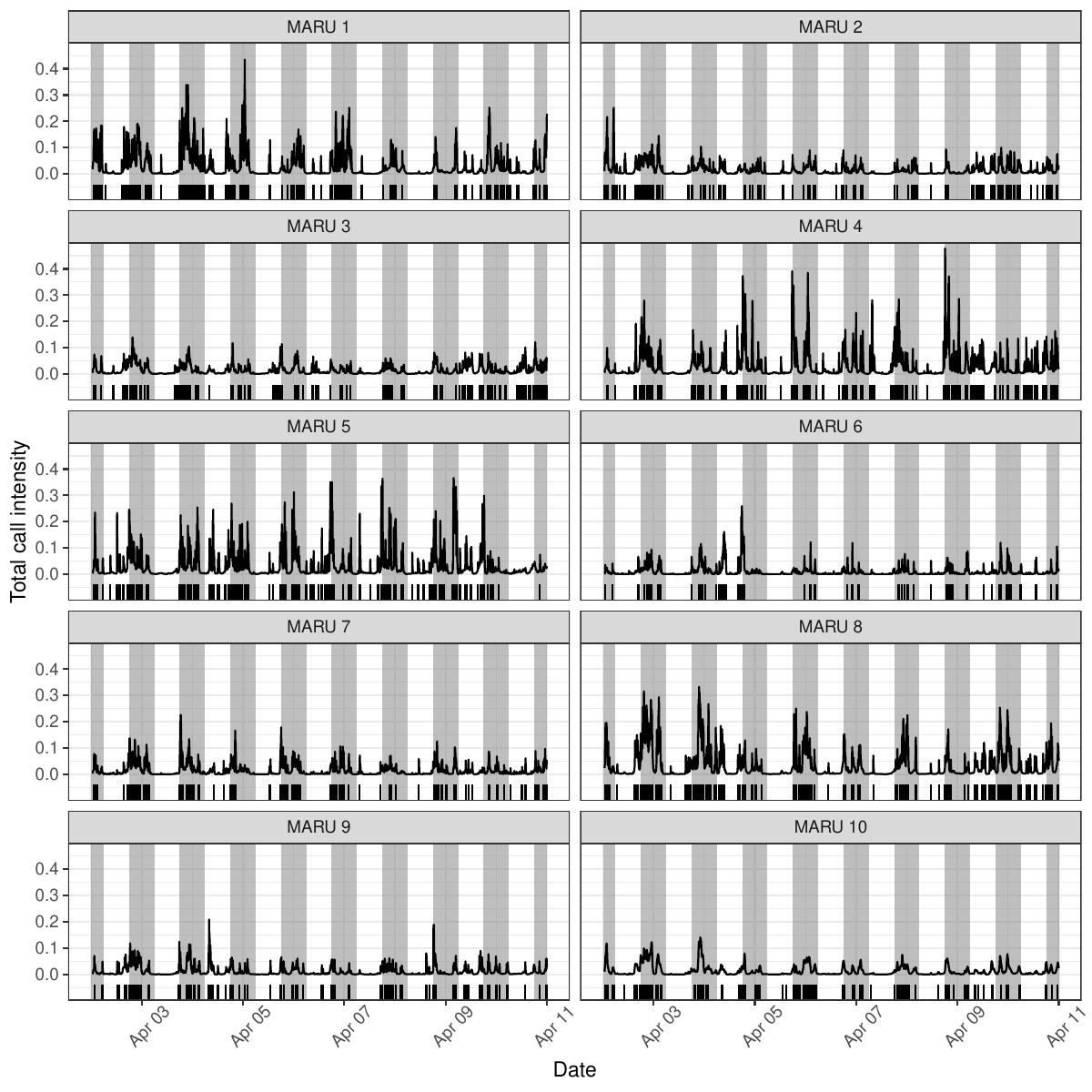}
\end{center}
\caption{Estimated intensity of total calls over time for each MARU. Rug tassels are the observed event sequence. Gray shade indicates 6 p.m. to 6 a.m.\label{sup:fig:ccbLamTotal}}
\end{figure}

In Figure S5 we show the estimated background intensity and countercall components across $t$ for each of the MARU's.  In Figure S6 we aggregate to show the overall intensity across $t$ for each MARU and overlay the event sequence for each MARU.

\bibliographystyle{apalike}
\bibliography{refs}